\documentclass{pspum-l}

\usepackage{amsmath,amssymb,amsfonts}
\usepackage{graphicx}
\usepackage{caption}
\usepackage{subcaption}

\newcommand{\CN}{\mathcal{N}}
\newcommand{\CS}{\mathcal{S}}
\newcommand{\CW}{\mathcal{W}}
\newcommand{\SU}{\mathrm{SU}}
\newcommand{\UU}{\mathrm{U}}
\newcommand{\tr}{\mathrm{tr}}
\newcommand{\dd}{\rm d}

\begin{document}

\title{2d SCFT from M-branes and its spectral network}

\author{Chan Y. Park}
\address{Mail Code 452-48\\California Institute of Technology\\Pasadena, CA 91125, USA}
\thanks{The author thanks Kentaro Hori and Yuji Tachikawa for the collaboration of the work \cite{Hori:2013ewa} on which this manuscript is based. The author also thanks Jihye Seo and Jaewon Song for carefully reading this manuscript and providing helpful comments.}
\subjclass[2010]{81T30}

\begin{abstract}
We consider the low-energy limit of the two-dimensional theory on multiple M2-branes suspended between a flat M5-brane and a curved M5-brane. We argue that it is described by an $\CN{=}(2,2)$ supersymmetric Landau-Ginzburg model with the superpotential determined by the shape of the curved M5-branes, which flows in the low-energy limit to a Kazama-Suzuki coset model. We provide evidence by studying ground states and BPS spectra of the systems. 
\end{abstract}

\maketitle

\section{Introduction}
Studying supersymmetric field theories as effective world-volume theories of branes has been a useful approach in understanding their non-perturbative aspects \cite{Hanany:1996ie,Hanany:1997vm}, especially when we analyze the brane configuration of such a theory with spectral network \cite{Gaiotto:2012rg}, which enables us to find the BPS spectra of a four-dimensional (4d) Seiberg-Witten theory or a two-dimensional (2d) theory in the 4d theory. Such approach has been applied in \cite{Hori:2013ewa} to provide evidence for an equivalence of 2d theories. Here we review a part of the paper, focusing on constructing relevant spectral networks and utilizing them to find BPS spectra of 2d theories. 

\section{Construction of spectral network}

\label{sec:SpectralNetwork}
Spectral network is introduced in \cite{Gaiotto:2012rg} as an extension of the analysis done in \cite{Klemm:1996bj,Shapere:1999xr}, building on the previous related work of \cite{Gaiotto:2009hg,Gaiotto:2010be,Gaiotto:2011tf}. Here we describe how to construct spectral networks of the 2d theories that we want to study. 

\subsection{$\CS$-wall} Consider $S[\mathfrak{g}]$, a 4d $\CN=2$ supersymmetric gauge theory of class $S$ \cite{Witten:1997sc, Gaiotto:2009we, Gaiotto:2009hg} associated to a Lie algebra $\mathfrak{g} = A_{K-1}$. A spectral network of $S[\mathfrak{g}]$ on the Coulomb branch consists of $\CS$-walls. Each $\CS$-wall carries two indices and follows the path described by the Seiberg-Witten curve and the differential of the 4d theory. When we have a Seiberg-Witten curve $f(t,x)=0$ as a multi-sheeted cover over the $t$-plane and the corresponding Seiberg-Witten differential $\lambda = x\, \dd t$, an $\CS_{jk}$-wall of a spectral network satisfies
\begin{align}
	\frac{\partial \lambda_{jk}}{\partial \tau} = \left( \lambda_j (t,x) - \lambda_k (t,x) \right) \frac{\dd t}{\dd \tau} = e^{i \theta}, \label{eq:S-wall}
\end{align}
where $\lambda_j$ is the value of $\lambda$ on the $j$-th sheet of $x$, and $\tau$ is a real parameter along the $\CS_{jk}$-wall.

\subsection{Spectral network around a branch point}
\begin{figure}[ht]
	\centering
	\begin{subfigure}{.3\textwidth}
		\centering
		\includegraphics[width=\textwidth]{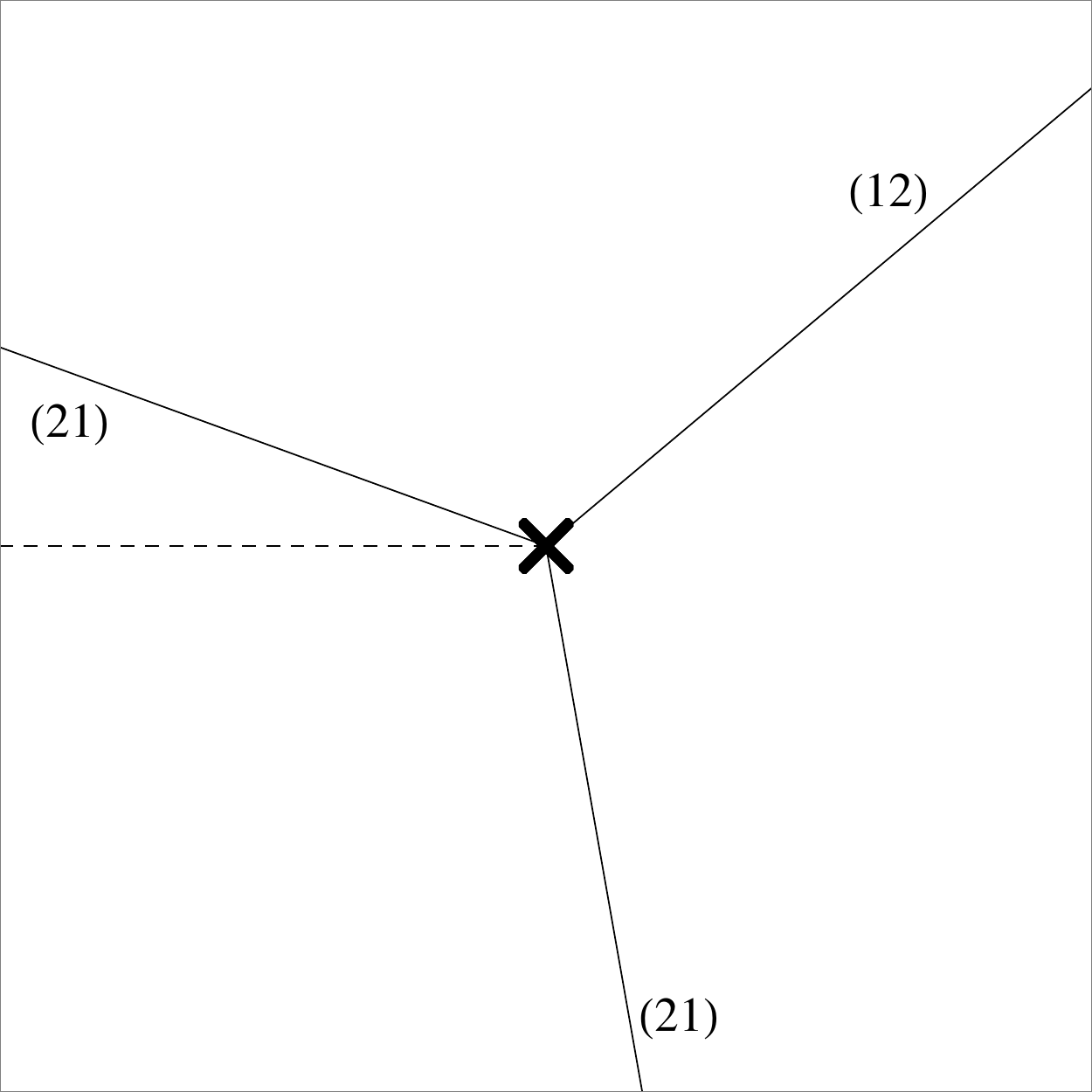}
		\caption{$N = 2$}
		\label{fig:bn2SN}
	\end{subfigure}
	\begin{subfigure}{.3\textwidth}
		\centering
		\includegraphics[width=\textwidth]{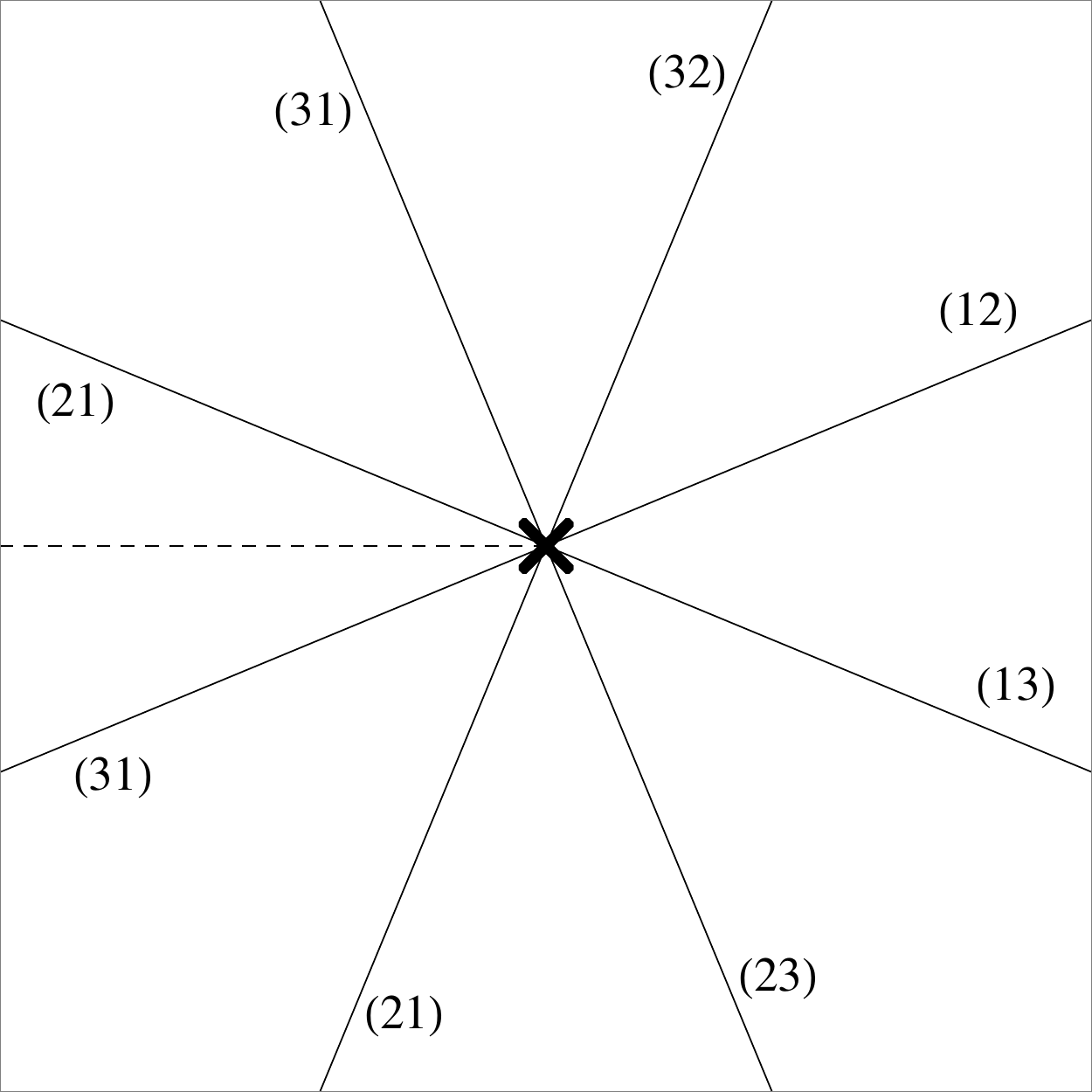}
		\caption{$N = 3$}
		\label{fig:bn3SN}
	\end{subfigure}
	\begin{subfigure}{.3\textwidth}
		\centering
		\includegraphics[width=\textwidth]{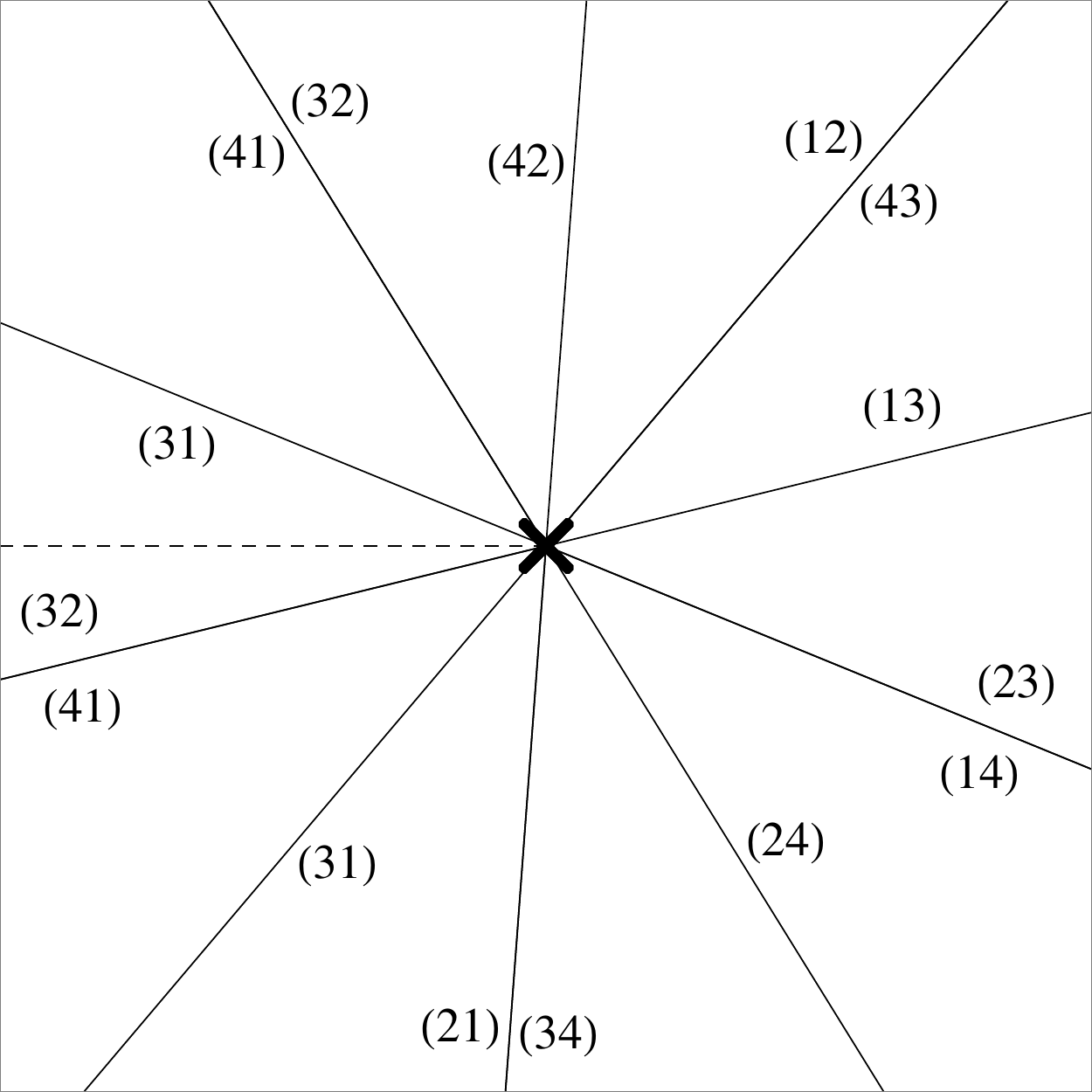}
		\caption{$N = 4$}
		\label{fig:bn4SN}
	\end{subfigure}
	\caption{$\CS$-walls around a branch point of ramification index $N$.
	         $\CS_{jk}$-walls are denoted by solid lines with $(jk)$. 
	         The broken line denotes the branch cut.}
	\label{fig:S-walls around a branch point}
\end{figure}
The spectral network from a branch point of ramification index $N$ consists of $\CS$-walls on the curve $t=x^N$ around $t=0$ with $\lambda = x\, \dd t$. On the $t$-plane, each $\CS_{jk}$-wall travels from the branch point along a real one-dimensional path defined by the differential equation (\ref{eq:S-wall}), which becomes in the neighborhood of $t=0$
\begin{align*}
	\omega_{jk} t^{1/N} \frac{\partial t}{\partial \tau} = \exp(i \theta),
\end{align*}
where 
\begin{align*}
	\omega_{jk} = \omega_j - \omega_k 
\end{align*}
and
\begin{align*}
	\omega_k = \exp \left( \frac{2\pi i}{N} k \right),\ k=0,1,\ \ldots,\ N-1.
\end{align*}
Then the solution for an $\mathcal{S}_{jk}$ is
\begin{align*}
	t_{jk}(\tau) =\left(\frac{\tau}{\omega_{jk}} \right)^{\frac{N}{N+1}} \exp \left( \frac{N}{N+1} i \theta \right)
\end{align*}
after rescaling $\tau$ to absorb a real numerical coefficient. The exponent $\frac{N}{N+1}$ makes the whole spectral network rotate by $\frac{2Nk\pi}{N+1}$ when we change $\theta$ from $0$ to $2k\pi$. Consistency of a spectral network under this rotation requires that there should be $N^{2}-1$ $\CS$-walls around $t=0$. The indices of $\CS$-walls are determined by choosing a branch cut. In \cite{Hori:2013ewa} it is described how to obtain such a spectral network starting from the spectral network of Figure \ref{fig:bn2SN}.

\begin{figure}[h]
	\centering
	\includegraphics[width=.3\textwidth]{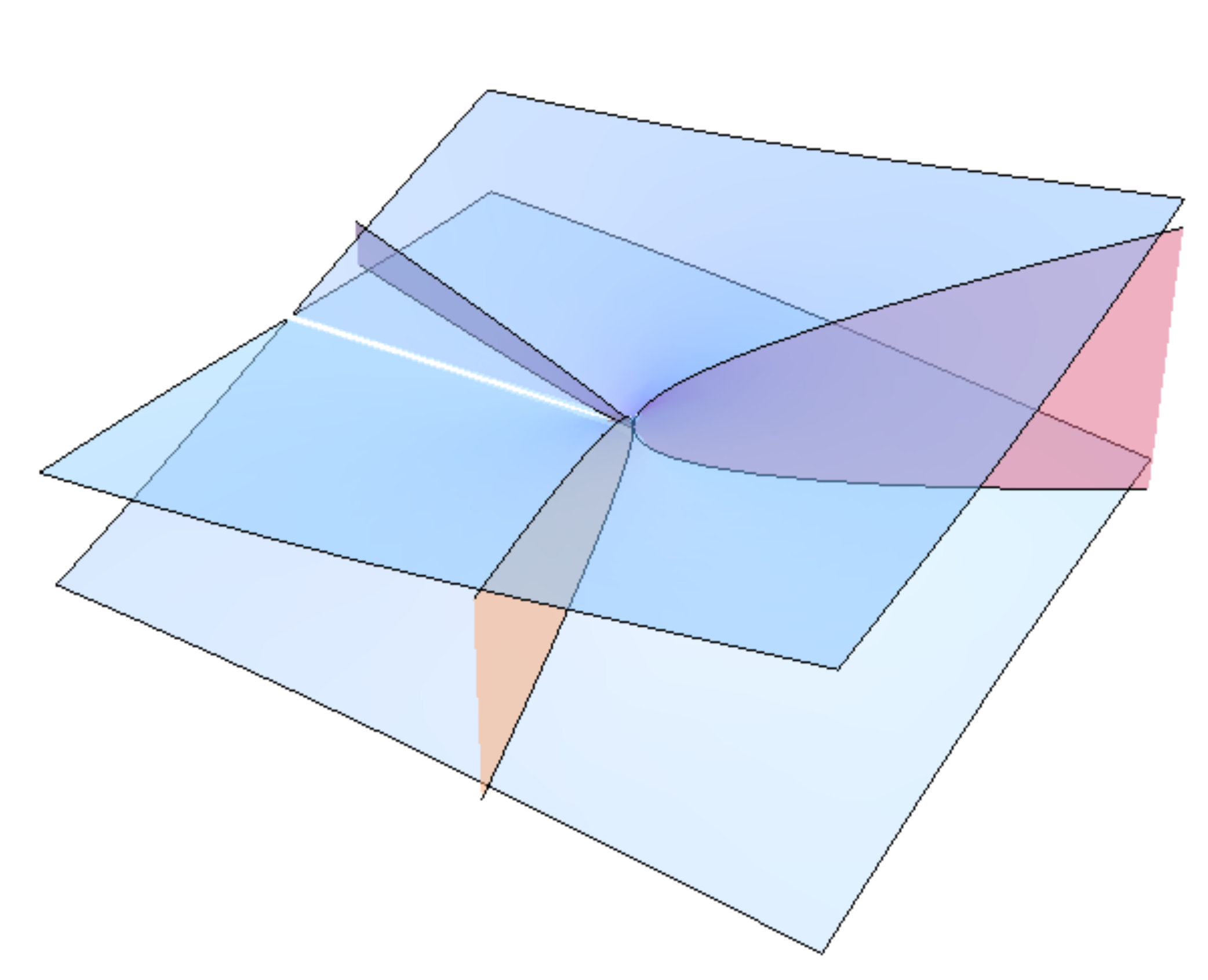}
	\caption{A Seiberg-Witten curve and $\CS$-walls around a branch point of index 2.}
	\label{fig:index_2_curve_and_S_wall}
\end{figure}
For a fixed $\theta$, each $\CS$-wall starts at the branch point and goes to infinity as shown in Figure \ref{fig:S-walls around a branch point}. Around a branch point of index 2, we have three $\mathcal{S}$-walls as shown in Figure \ref{fig:bn2SN}. Figure \ref{fig:index_2_curve_and_S_wall} illustrates the corresponding Seiberg-Witten curve by plotting the real part of $x$ over the $t$-plane, and also real two-dimensional surfaces that end on the curve along the $\CS$-walls are shown. Figures \ref{fig:bn3SN} and \ref{fig:bn4SN} show spectral networks around a branch point of index 3 and 4, respectively.

\section{2d $\CN=(2,2)$ theories from M2-branes}
Here we show the brane configuration that describes the 2d theory of our interest, make the main claim of the equivalence of the 2d theory and a Landau-Ginzburg model, and provide evidence for the claim.

\subsection{M2-branes ending near a ramification point of M5-branes}

\begin{figure}[h]
	\centering
	\begin{subfigure}{.6\textwidth}
		\centering
		\begin{tabular}{l||c|c|c|c|c|c|c|c|c|c}
			& $x^0$ & $x^1$ & $x^2$ & $x^3$ & $v$ & $x^6$ & $t$ & $x^8$& $x^9$ \\
			\hline 
			{M5} & $-$ & $-$ & $-$ & $-$ & $-$ & $L$ & $t(v)$ & $\cdot$ & $\cdot$  \\
			{M5$'$}& $-$ & $-$ & $\cdot$ & $\cdot$ & $-$ & $\cdot$ & $t_0$ & $-$ & $-$ \\
			{M2}& $-$ & $-$ & $\cdot$ & $\cdot$ & $\sigma$ & $-$ & $t_0$ & $\cdot$ & $\cdot$
		\end{tabular}
	\end{subfigure}
	\begin{subfigure}{.375\textwidth}
		\centering
		\includegraphics[width=\textwidth]{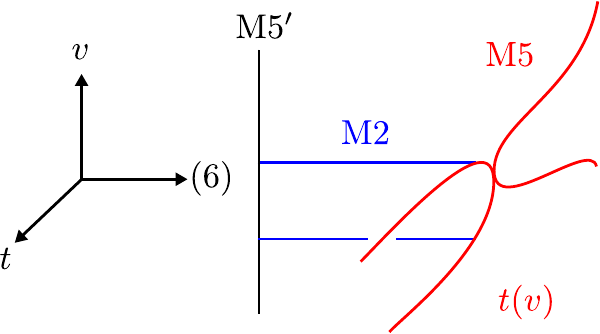}
	\end{subfigure}
	\caption{Configuration of branes. $v=x^4+ix^5$ and $t=\exp(x^7+ix^{10})$.}
	\label{fig:M_brane_2D_N_2_2}
\end{figure}

Consider the brane configuration of Figure \ref{fig:M_brane_2D_N_2_2}. There is an M5-brane wrapping a curve $t(v)$, which is an $N$-sheeted cover over the $t$-plane, and $k$ M2-branes ending on the curve. The other ends of the M2-branes are on a flat M5-brane, which we call M5$'$. The s-rule \cite{Hanany:1996ie, Hanany:1997vm} applies to this brane configuration, and there is only one M2-brane connecting one sheet of the curved M5-brane and M5$'$. 

The brane configuration preserves four supercharges, resulting in a 2d $\CN=(2,2)$ theory from M2-branes that spans $1+1$-dimensional subspace of the 4d spacetime \cite{Hanany:1997vm}. A set of flat M2-branes connecting the two M5-branes gives a ground state of the 2d theory, and a configuration of M2-branes interpolating two ground state M2-branes gives a 2d solitonic BPS state \cite{Hanany:1997vm, Dorey:1998yh}, which becomes massless when the endpoint of M2-branes is on a ramification point \cite{Gaiotto:2011tf}.

\subsection{2d BPS states from a spectral network}
From the spectral network of the curved M5-brane we can read out the BPS spectrum of the 2d theory from an M2-brane ending at a point on the Seiberg-Witten curve $t(v)$. A flat M2-brane ending at $(t,v)=(t_0,v_j)$, where
\begin{align*}
	t(v_j) = t_0,\ j = 1,\ldots,N,
\end{align*}
gives the $j$-th ground state of the 2d theory. When an $\CS_{jk}$-wall connects a $(jk)$-branch point to the endpoint $t=t_0$, it corresponds to a BPS soliton interpolating two 2d ground states corresponding to $(t_0, v_j)$ and $(t_0, v_k)$ \cite{Gaiotto:2011tf}. The central charge of the BPS state is calculated by integrating $\lambda_{jk}$ along the finite $\CS$-wall,
\begin{align}
	Z = \int_{\tau_b}^{\tau_s} \lambda_{jk}(t) \frac{\partial t}{\partial \tau} d\tau = \int_{\tau_b}^{\tau_s} e^{i\theta} d\tau, \label{eq:Z_2d}
\end{align}
where $t(\tau_b)$ is the branch point and $t(\tau_s) = t_0$.

\begin{figure}[ht]
	\centering
	\begin{subfigure}{.22\textwidth}
		\centering
		\includegraphics[width=\textwidth]{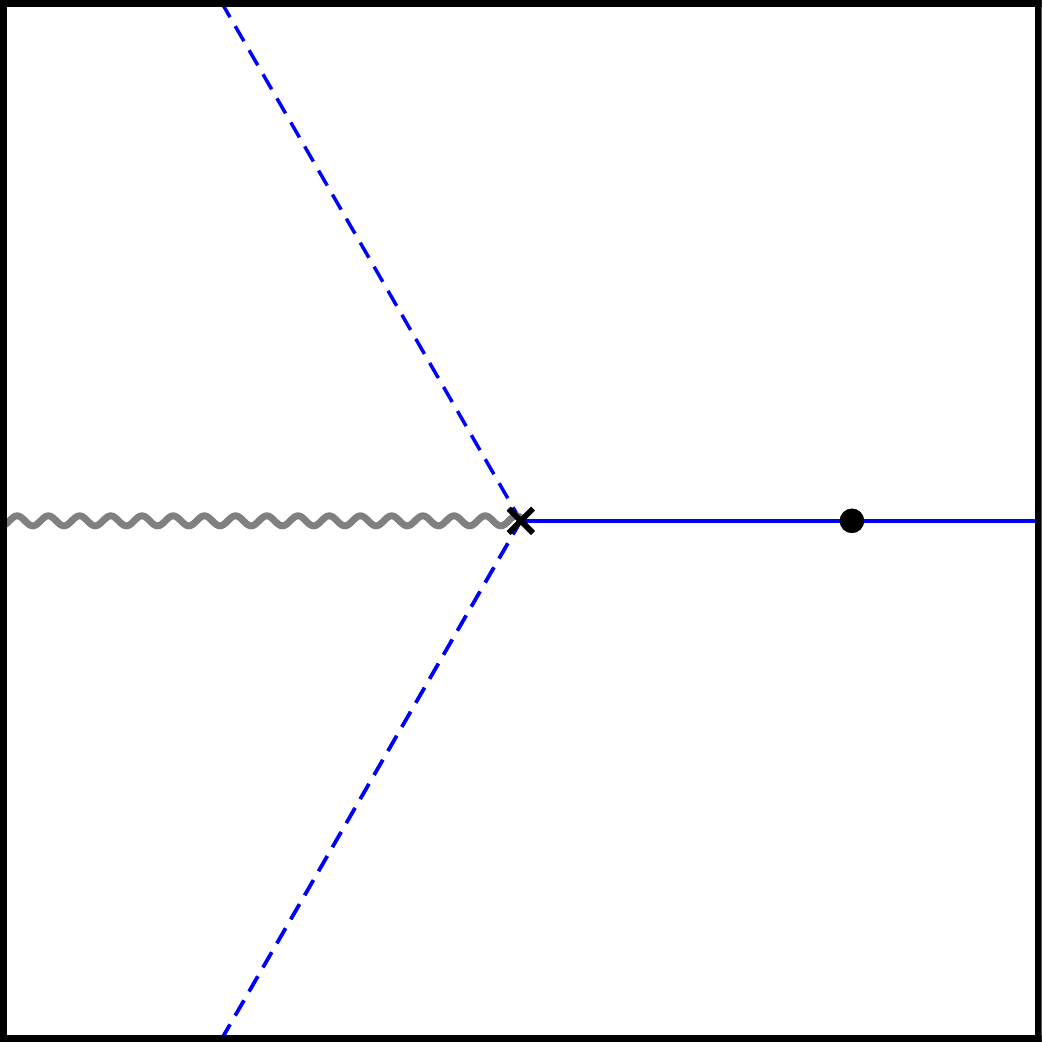}
		\caption{$\theta = 0$}
		\label{fig:bn2SN_01}
	\end{subfigure}
\hspace{5pt}	
	\begin{subfigure}{.22\textwidth}
		\centering
		\includegraphics[width=\textwidth]{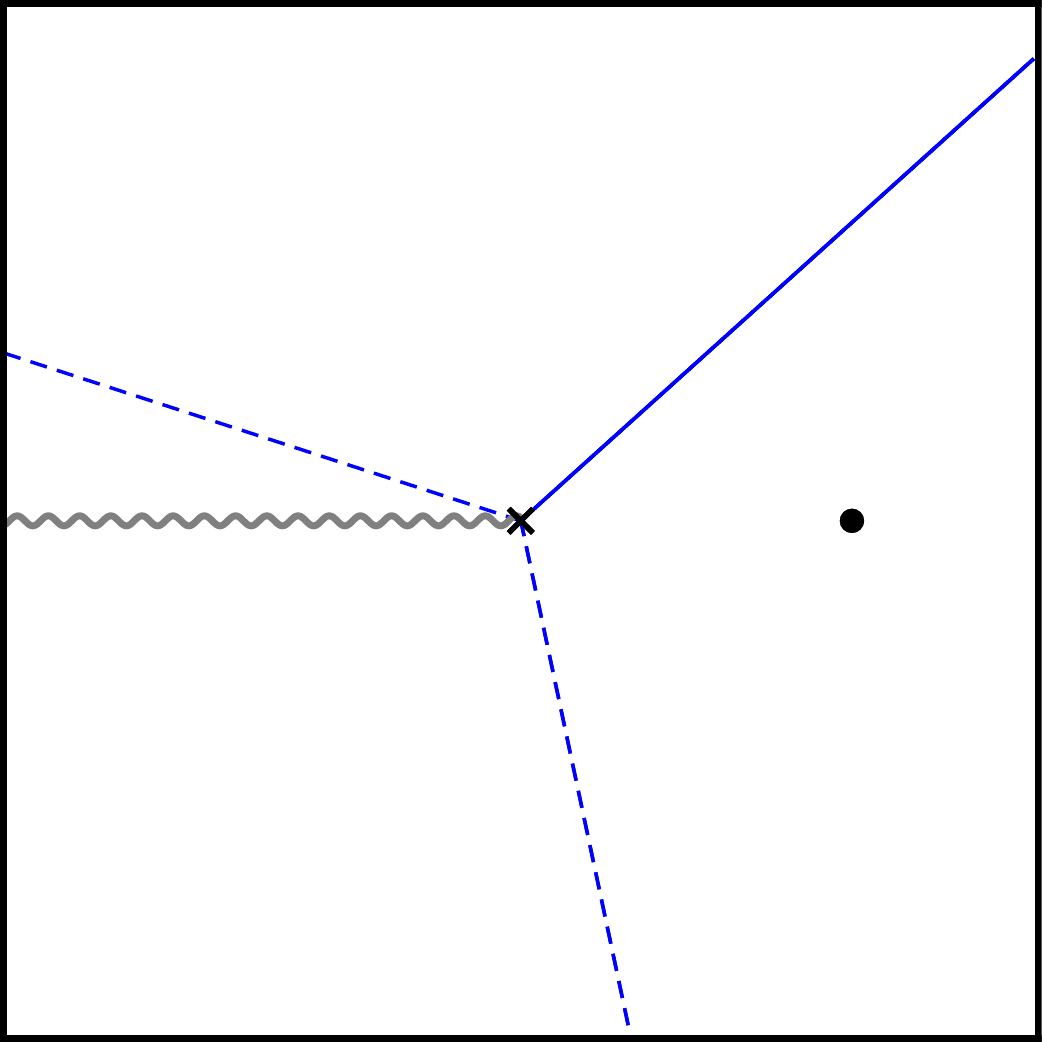}
		\caption{$\theta \approx \pi/3$}
		\label{fig:bn2SN_02}
	\end{subfigure}
\hspace{5pt}	
	\begin{subfigure}{.22\textwidth}
		\centering
		\includegraphics[width=\textwidth]{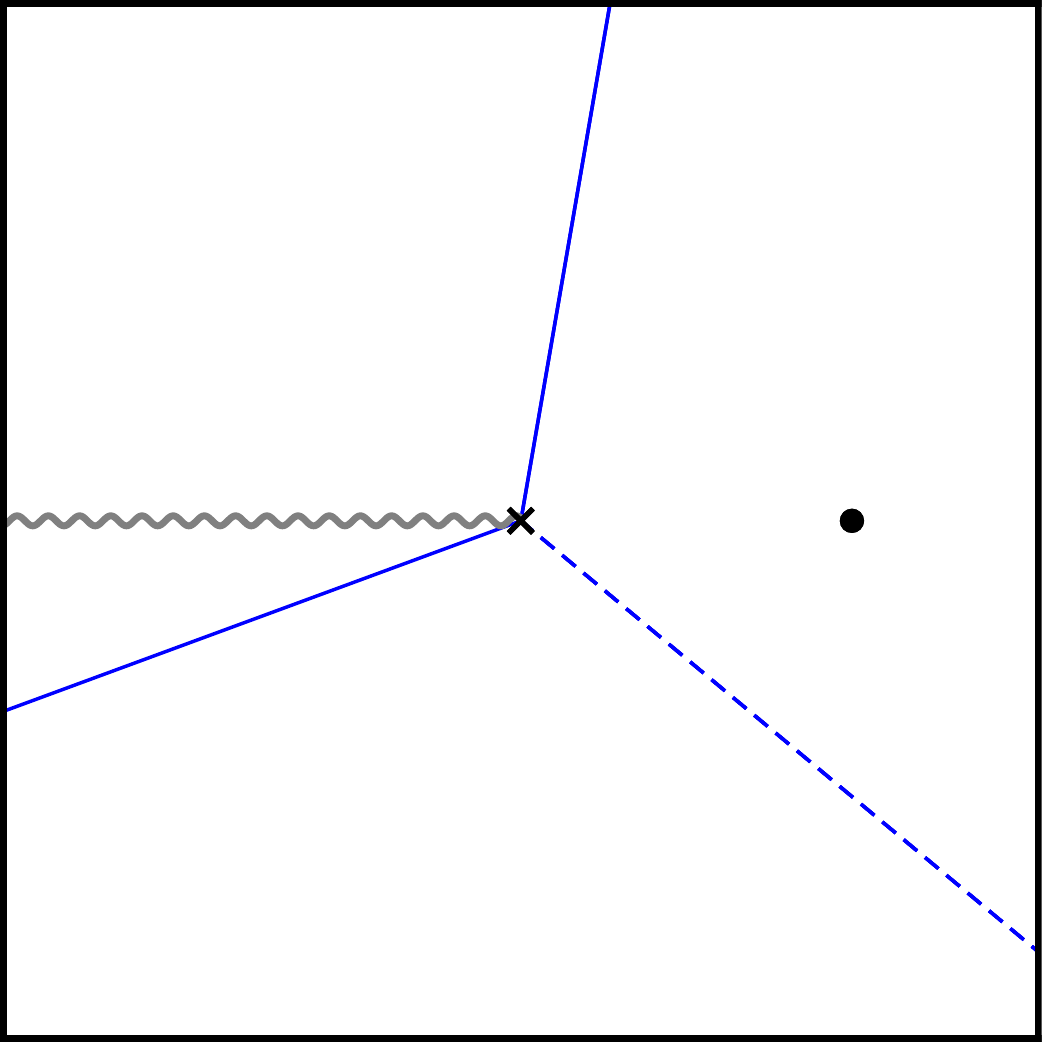}
		\caption{$\theta \approx 2\pi/3$}
		\label{fig:bn2SN_03}
	\end{subfigure}
\hspace{5pt}
	\begin{subfigure}{.22\textwidth}
		\centering
		\includegraphics[width=\textwidth]{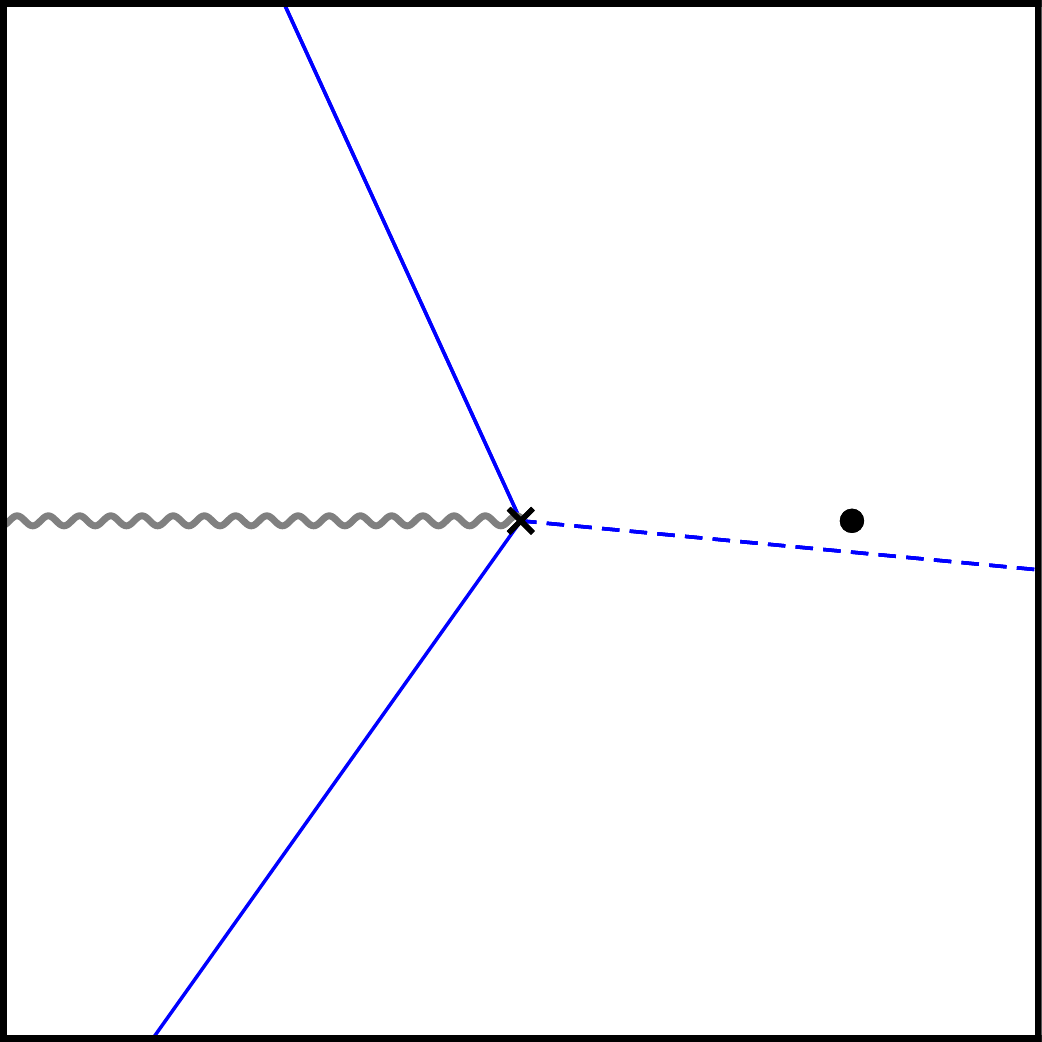}
		\caption{$\theta \approx \pi$}
		\label{fig:bn2SN_04}
	\end{subfigure}
	
	\caption{Spectral network around a branch point of index 2 at various values of $\theta$, $0 \leq \theta <\pi$, and the endpoint of an M2-brane.}
	\label{fig:bn2SN_theta}
\end{figure}

To find out all the BPS states, we examine which $\CS$-walls pass the M2-brane endpoint as we change $\theta$. When such an $\CS$-wall exists, the value of $\theta$ is the phase of the central charge of the corresponding BPS state. See Figure \ref{fig:bn2SN_theta}, where we have two 2d BPS states with $\arg(Z_{12}) = 0$ and $\arg(Z_{21})=\pi$.

\subsection{2d $\mathcal{N}=(2,2)$ theory from multiple M2-branes}
In \cite{Hori:2013ewa} it is argued that the 2d theory {$M(N,k;\mu_j)$} from the $k$ M2-branes between M5$'$ and the curved M5-brane wrapping a Seiberg-Witten curve
\begin{align*}
	t - t_0 = \partial_v {P(v)} = v^N + \mu_2 v^{N-2} + \cdots + \mu_N
\end{align*}
is described by a 2d $\CN=(2,2)$ Landau-Ginzburg model {$LG(N,k;\mu_j)$} with chiral fields $X_1,\ldots,X_k$ and superpotential $W=W(X_1,\ldots,X_k)$ whose functional form is defined as
\begin{align*}
	W(s_1,\ldots,s_k)=\tr {P(\Phi)},
\end{align*} 
where $s_j$ is an elementary symmetric polynomial of degree $j$ in $k$ auxiliary variables $\phi_j$,
\begin{align*}
	\Phi = \mathrm{diag}(\phi_1,\ldots,\phi_k),\ \prod_{a=1}^k (z-\phi_a) = z^k + s_1 z^{k-1} + \cdots + s_k.
\end{align*}

For {$M(N,k;\mu_j)$}, $\mu_j$ determine the locations of $(N-1)$ number of index 2 branch points of the curve on the $t$-plane. When all $\mu_j$ vanish, the branch points collide at $t=t_0$ and form an index $N$ branch point, corresponding to having M2-branes ending on the ramification point of the curved M5-brane.	We will call the resulting 2d theory $M(N,k) \equiv M(N,k;\mu_j=0)$. For the Landau-Ginzburg model, setting all $\mu_j$ to zero corresponds to flowing the theory to its infra-red (IR) fixed point, which we will call $LG(N,k) \equiv LG(N,k;\mu_j=0)$. This is claimed in \cite{Gepner:1988wi,Lerche:1989uy} to be described by an $N=2$ superconformal coset model 
\begin{align*}
	\frac{G_k}{H} = \frac{\SU(N)_1}{\mathrm{S} \left[\UU(k) \times \UU(N-k) \right]}, 
\end{align*}
constructed by Kazama and Suzuki \cite{Kazama:1988qp}, which we will call {$KS(N,k)$}. Then $\mu_j$ can be considered as giving the relevant perturbations 
\begin{align*}
	\tr \left[ \mu_j  \Phi^{N+1-j} \right]
\end{align*}
of $LG(N,k)$ from $KS(N,k)$ with scaling dimension 
\begin{align*}
	\Delta(\mu_j) = \frac{j}{N+1}.
\end{align*}

To summarize, we claim that both $M(N,k)$, the 2d theory from the brane configuration, and $LG(N,k)$, the Landau-Ginzburg model with a superpotential, flow in the IR to the same fixed point described by a coset model $KS(N,k)$. When $k=1$, $M(N,1)$ is argued in \cite{Tong:2006pa} to be described by the IR limit of a Landau-Ginzburg model with superpotential 
\begin{align*}
	W(X) = \frac{X^{N+1}}{N+1}, 
\end{align*}
which is the $A_{N-1}$ minimal model that corresponds to a coset model $KS(N,1)$. This is consistent with the claim.

\subsection{Evidence for the claim}
The 2d theories have the same Witten index, or the number of ground states. The number of ground states of $M(N,k;\mu_j)$ is $\binom{N}{k} = \frac{N!}{k!(N-k)!}$ due to the s-rule \cite{Hanany:1996ie,Hanany:1997vm}, and we expect the Witten index to be maintained as we take $\mu_j \to 0$. The same is true for $LG(N,k;\mu_j)$. The Witten index of $KS(N,k)$ is also $\binom{N}{k}$ \cite{Lerche:1989uy}. 
		
The 2d theories enjoy the $k \leftrightarrow (N-k)$ duality. For $M(N,k;\mu_j)$ the duality comes from the Hanany-Witten transition \cite{Hanany:1996ie}: when we move M5$'$ across the curved M5-brane, $k$ M2-branes disappear and $(N-k)$ M2-branes are created. For $LG(N,k;\mu_j)$, when $k > (N-k)$ we can re-express everything in terms of $(N-k)$ variables. The coset model trivially has the same duality because of the form of the coset, $KS(N,k) = KS(N,N-k)$.

In addition to these basic checks, we will see that the BPS spectrum of 
\begin{align*}
	M'(N,k,\mu_N) \equiv M(N,k;\mu_2 = \cdots = \mu_{N-1} = 0 \neq \mu_N)
\end{align*}
matches with the BPS spectrum of $LG'(N,k,\mu_N)$, which is the deformation from the IR fixed point $KS(N,k)$ with the most relevant term $\mu_{N} X_1$ \cite{Fendley:1990us,Fendley:1990zj,Lerche:1991re}. Here we summarize the BPS spectrum of the brane configuration we obtain via spectral network in Section \ref{sec:2d BPS spectrum from spectral network}.

\begin{figure}[h]
	\begin{subfigure}[b]{.2\textwidth}		
		\includegraphics[width=\textwidth]{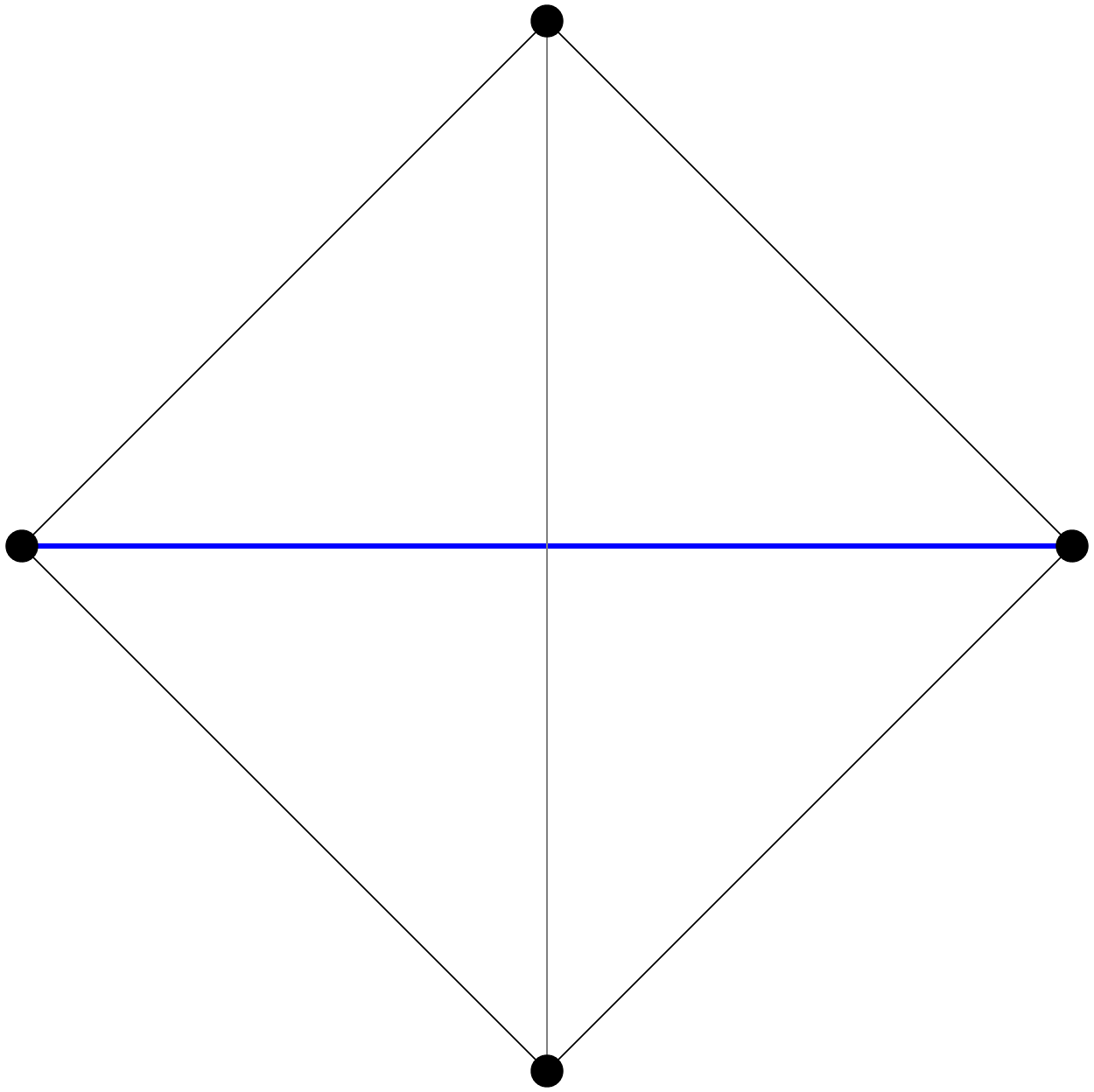}
		\caption{$N=4$, $k=1$}
	\end{subfigure}
	\hspace{1em}
	\begin{subfigure}[b]{.2\textwidth}		
		\includegraphics[width=\textwidth]{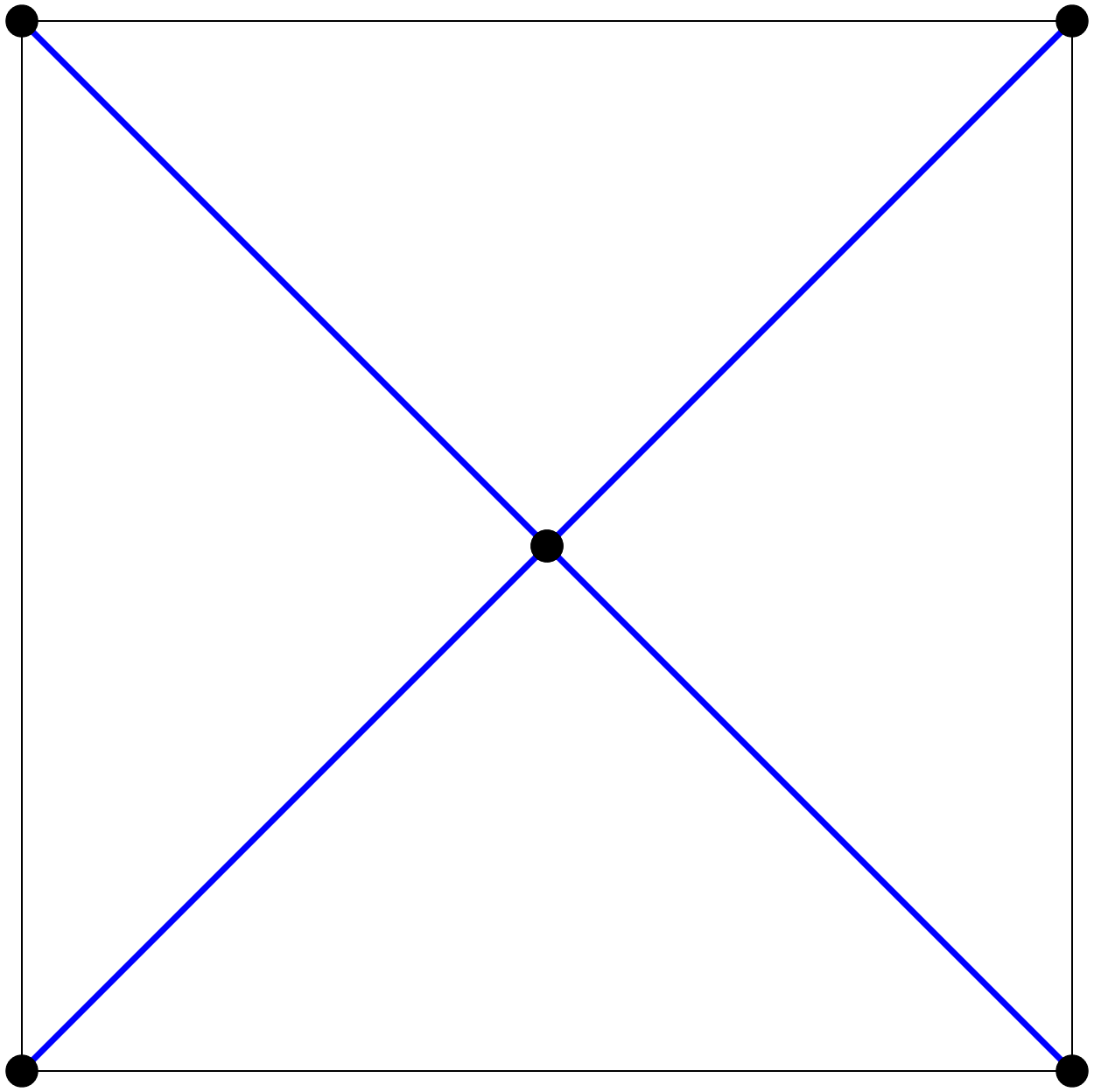}
		\caption{$N=4$, $k=2$}
	\end{subfigure}
	\hspace{1em}
	\begin{subfigure}[b]{.2\textwidth}		
		\includegraphics[width=\textwidth]{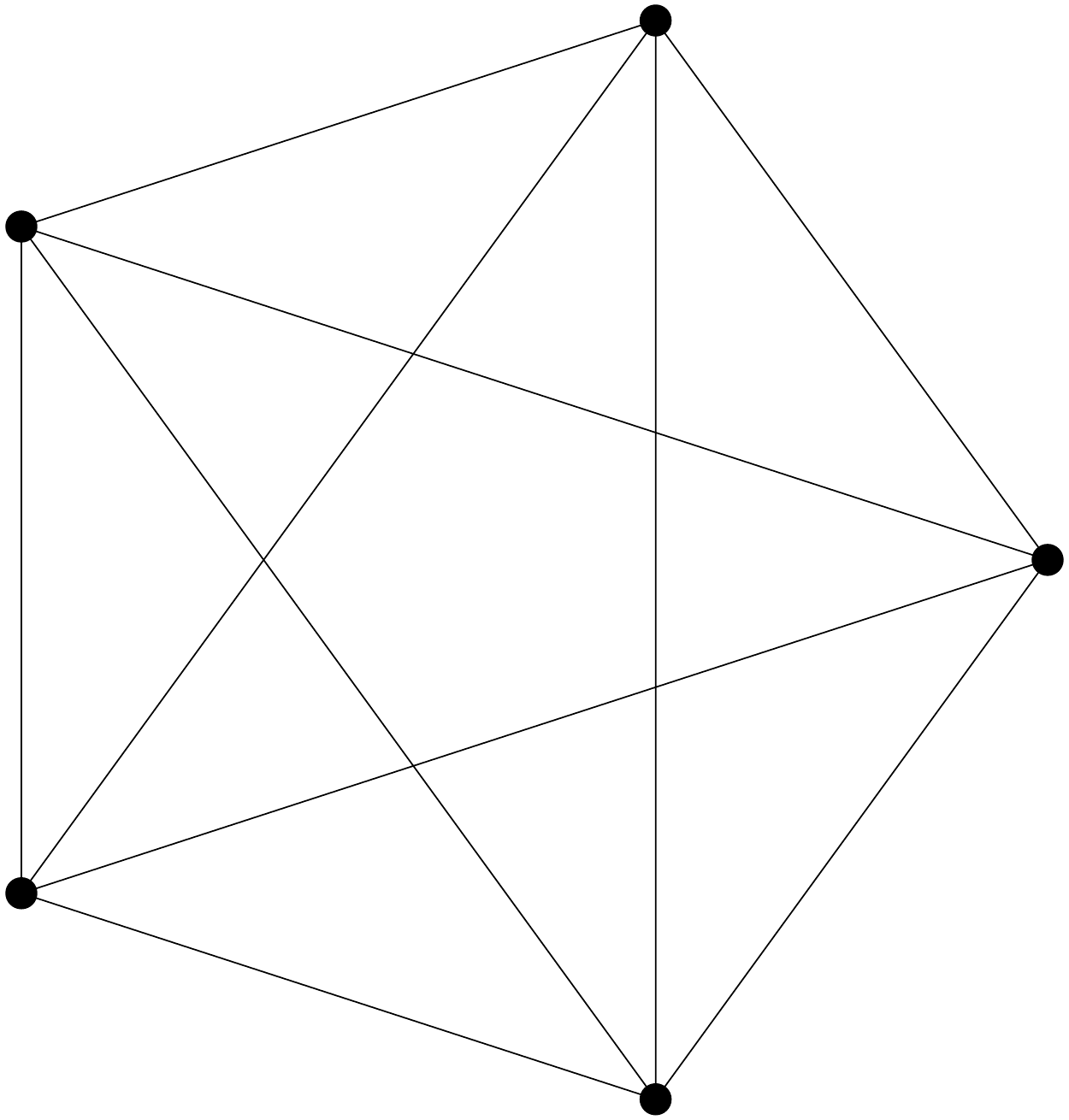}
		\caption{$N=5$, $k=1$}
	\end{subfigure}		
	\caption{Ground states and solitons on the $W$-plane}
	\label{fig:W_plane_projection}
\end{figure}

The ground states of $M'(N,k,\mu_N)$ can be identified with the weights of $k$-th exterior power of the fundamental representation of $SU(N)$, and the solitons of the theory are given by roots connecting two weights in the representation. To obtain the $W$-plane description of them, we project the weight space onto a complex plane such that all the weights of the fundamental representation, which form the vertices of an $N$-simplex, are shown on a circle so that the roots form a Petrie polygon, see Figure \ref{fig:W_plane_projection} for examples.
The same structure describes the ground states and the BPS spectra of $LG'(N,k,\mu_N)$ \cite{Lerche:1991re}, providing a good piece of evidence for the claim.

\section{2d BPS spectrum from spectral network}
\label{sec:2d BPS spectrum from spectral network}
Here we illustrate with examples how to obtain the BPS spectrum of $M(N,k;\mu_j)$ using spectral network.

\subsection{$M'(N=3,k=1,\mu_3)$}
\begin{figure}[ht]
	\begin{subfigure}[b]{.3\textwidth}
		\centering
		\includegraphics[width=\textwidth]{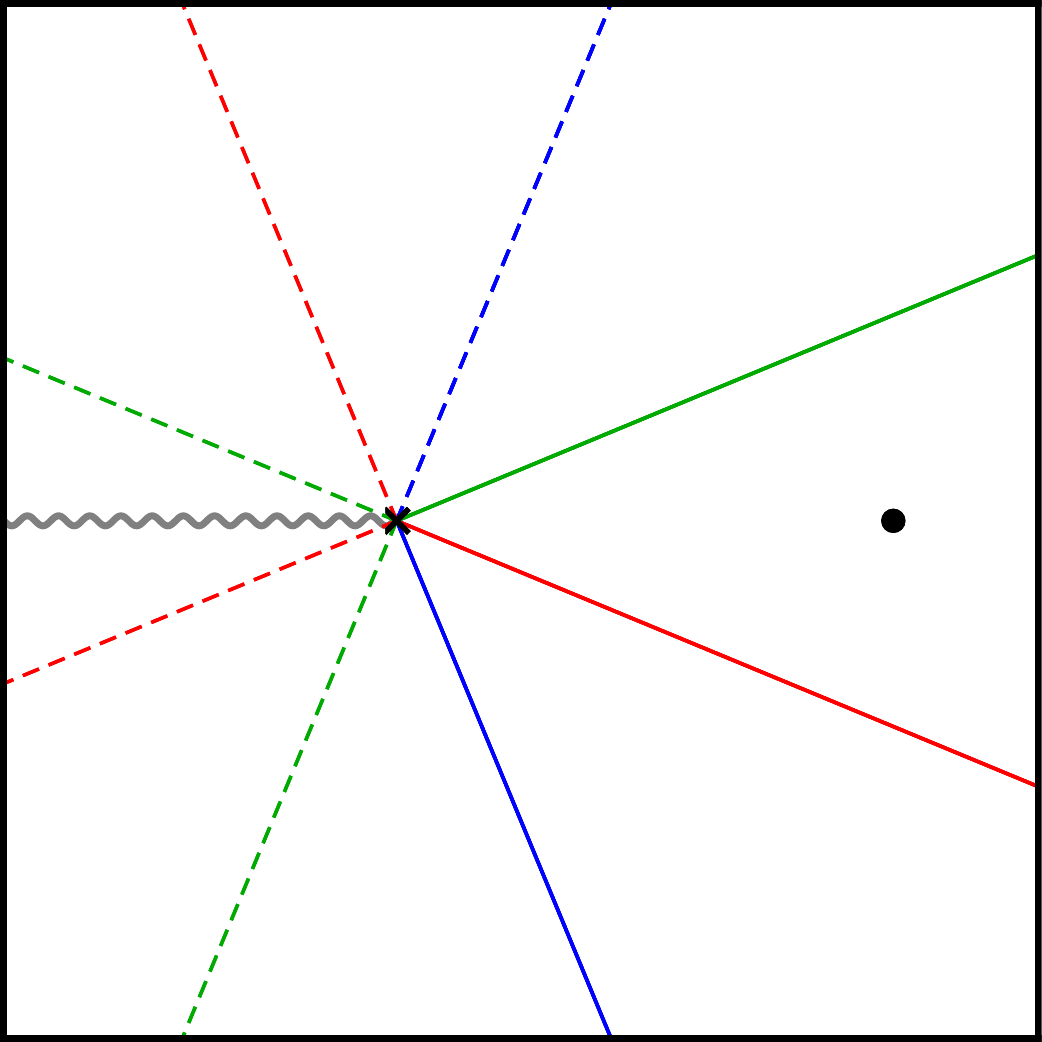}
		\vspace{0em}
		\caption{spectral network}
		\label{fig:IRMN3k_SWall_mu2_zero}
	\end{subfigure}
	\begin{subfigure}[b]{.6\textwidth}
		\centering
		\includegraphics[width=.5\textwidth]{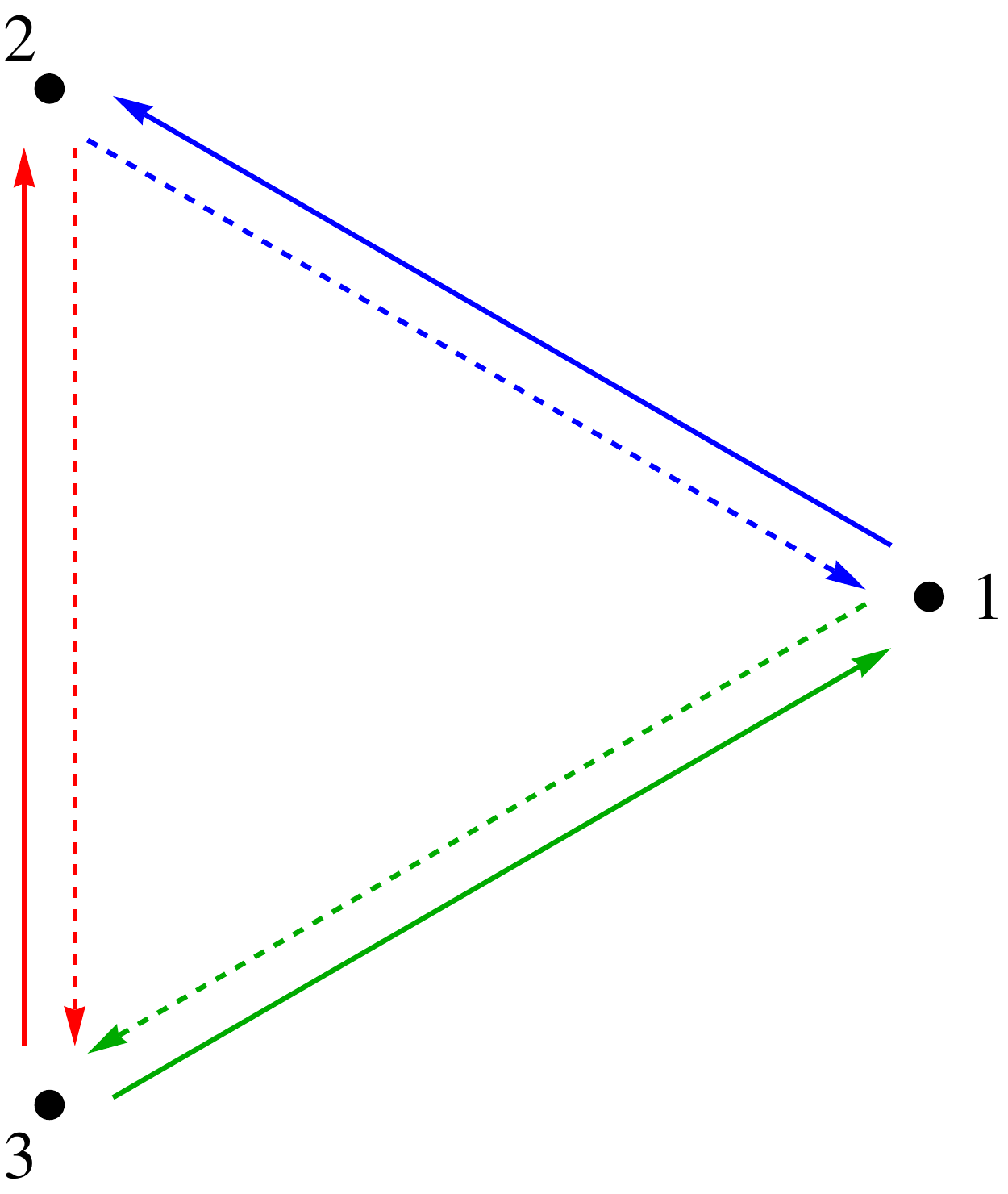}
		\caption{Ground states and solitons on the $v$-plane}
	\label{fig:IRMN3k1_v_plane}
	\end{subfigure}
	\caption{$M'(N=3,k=1,\mu_3)$}
\end{figure}

Thanks to the symmetric configuration of the spectral network as shown in Figure \ref{fig:IRMN3k_SWall_mu2_zero}, we can analytically find the BPS spectrum of the theory using (\ref{eq:Z_2d}). The ground states and the central charges of the BPS states can be represented as shown in Figure \ref{fig:IRMN3k1_v_plane}, where a dot with a number $i$ represents the value of $v$ of the $i$-th ground state M2-brane, and the central charge of the $ij$-soliton is shown as an arrow from $j$ to $i$ because it is proportional to the difference of $v_i$ and $v_j$. 

This can also be read as ground states and solitons on the $W$-plane of a Landau-Ginzburg model with three ground states, which is consistent with the equivalence between $M'(N,k,\mu_3)$ and $LG'(N,k,\mu_3)$. We can identify the ground states as the fundamental representation of $A_2$, and the solitons with the roots connecting the weights of the representation.

\subsection{$M'(N=3,k=2,\mu_3)$}

\begin{figure}[ht]
	\begin{subfigure}[b]{.6\textwidth}		
		\centering
		\includegraphics[width=\textwidth]{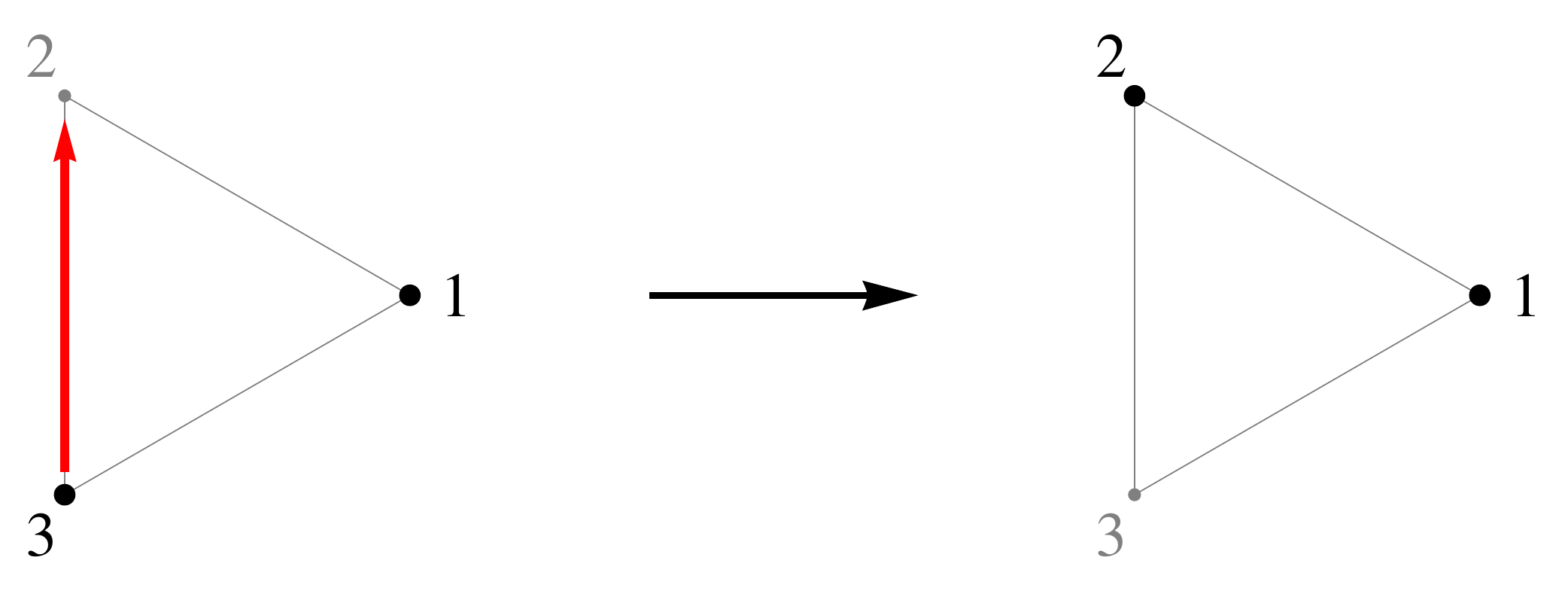}
		\vspace{-5pt}
		\caption{$[13] \to [12]$ soliton}
		\label{fig:IRMN3k2_soliton}
	\end{subfigure}
	\hspace{1em}
	\begin{subfigure}[b]{.3\textwidth}		
		\centering
		\includegraphics[width=\textwidth]{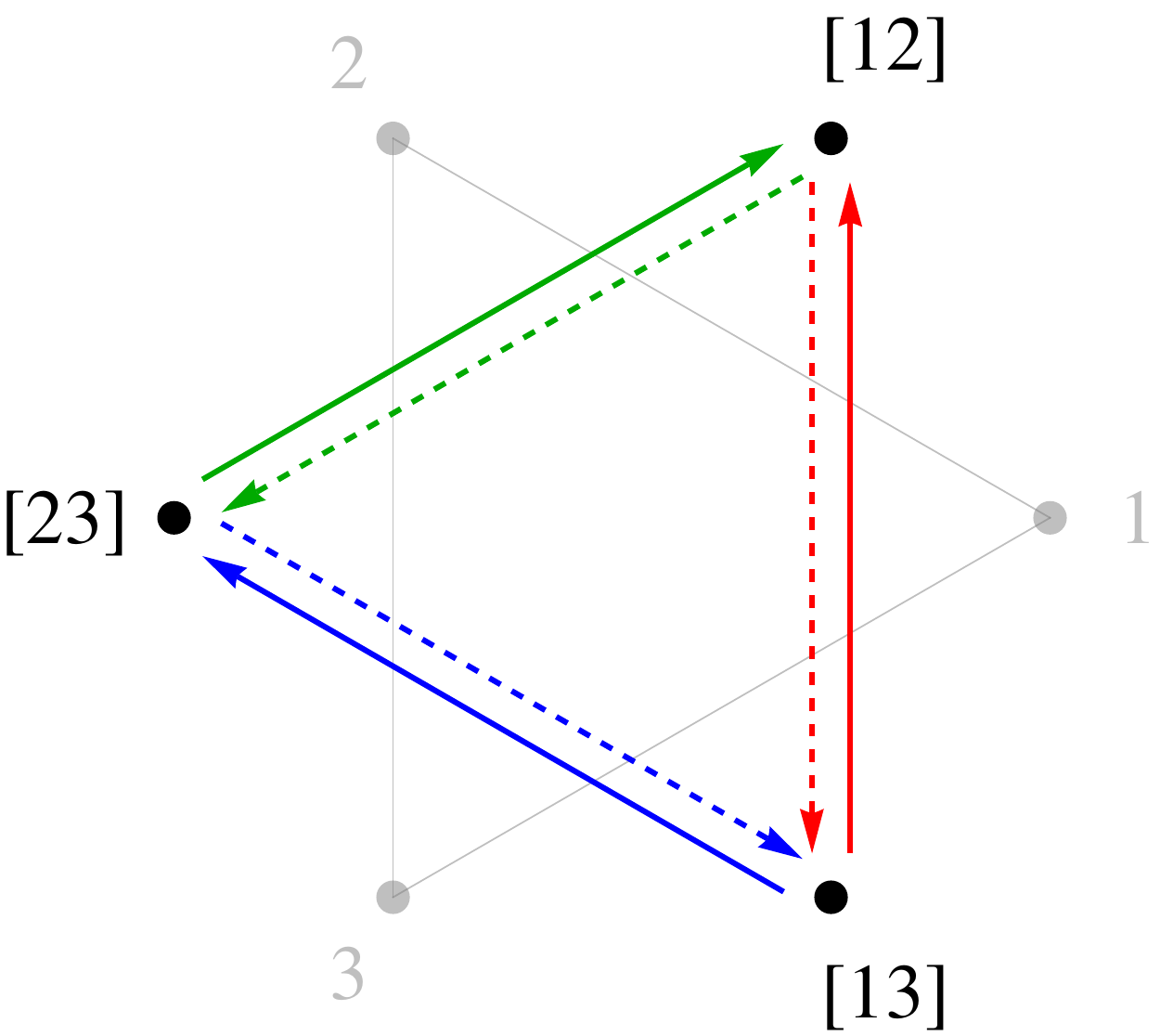}
		\caption{on the $v$-plane}
		\label{fig:IRMN3k2_v_plane}
	\end{subfigure}
	\caption{ground states and solitons of $M'(N=3,k=2,\mu_3)$}
	\label{fig:IRMN3k2}
\end{figure}
	
For the 2d theory from two M2-branes ending on the M5-brane, a ground state is two M2-branes occupying two among the three possible places, as shown in Figure \ref{fig:IRMN3k2_soliton}, where we denote a ground state of two M2-branes occupying $j$-th \& $k$-th $k=1$ ground states as $[jk]$. This is because the s-rule prevents two M2-branes being at the same location of $v$. Then a soliton interpolating two different ground states can be obtained from the $k=1$ BPS spectrum. For example, Figure \ref{fig:IRMN3k2_soliton} shows a $k=2$ soliton interpolating $[13]$ and $[12]$ ground states, which is from the $k=1$ soliton interpolating the third and the second ground states.

We can again arrange the ground states and the solitons on the $v$-plane as shown in Figure \ref{fig:IRMN3k2_v_plane}, which can be considered as the $W$-plane diagram of the Landau-Ginzburg model. Here we see that the duality of $k \leftrightarrow (N-k)$ corresponds to the complex conjugation of the representation of the ground states.

\subsection{$M'(N=4,k=1,\mu_4)$}
\begin{figure}[h]
	\begin{subfigure}[b]{.35\textwidth}		
		\centering
		\includegraphics[width=\textwidth]{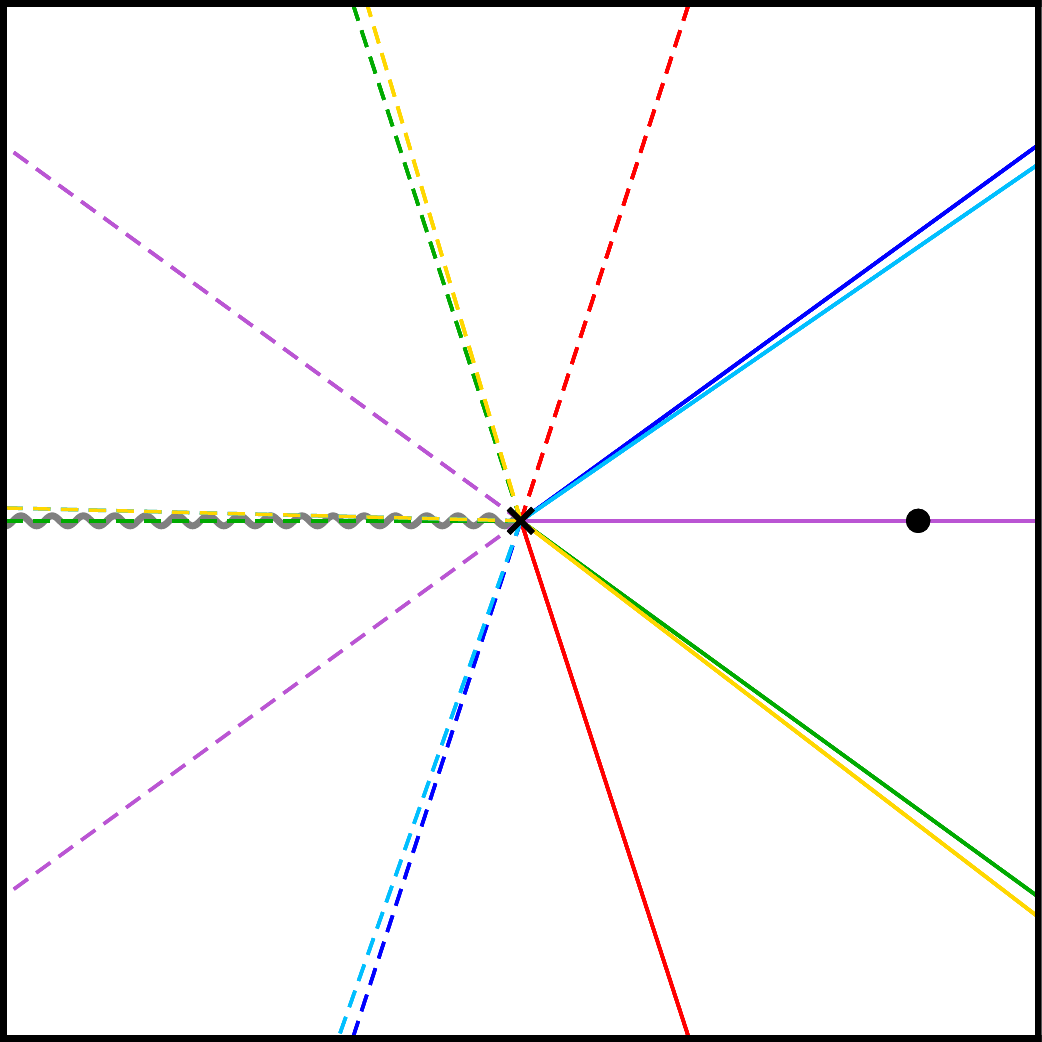}
		\vspace{0pt}
		\caption{spectral network}
		\label{fig:IRMN4k1_SWall}
	\end{subfigure}
	\hspace{.5em}
	\begin{subfigure}[b]{.3\textwidth}		
		\includegraphics[width=\textwidth]{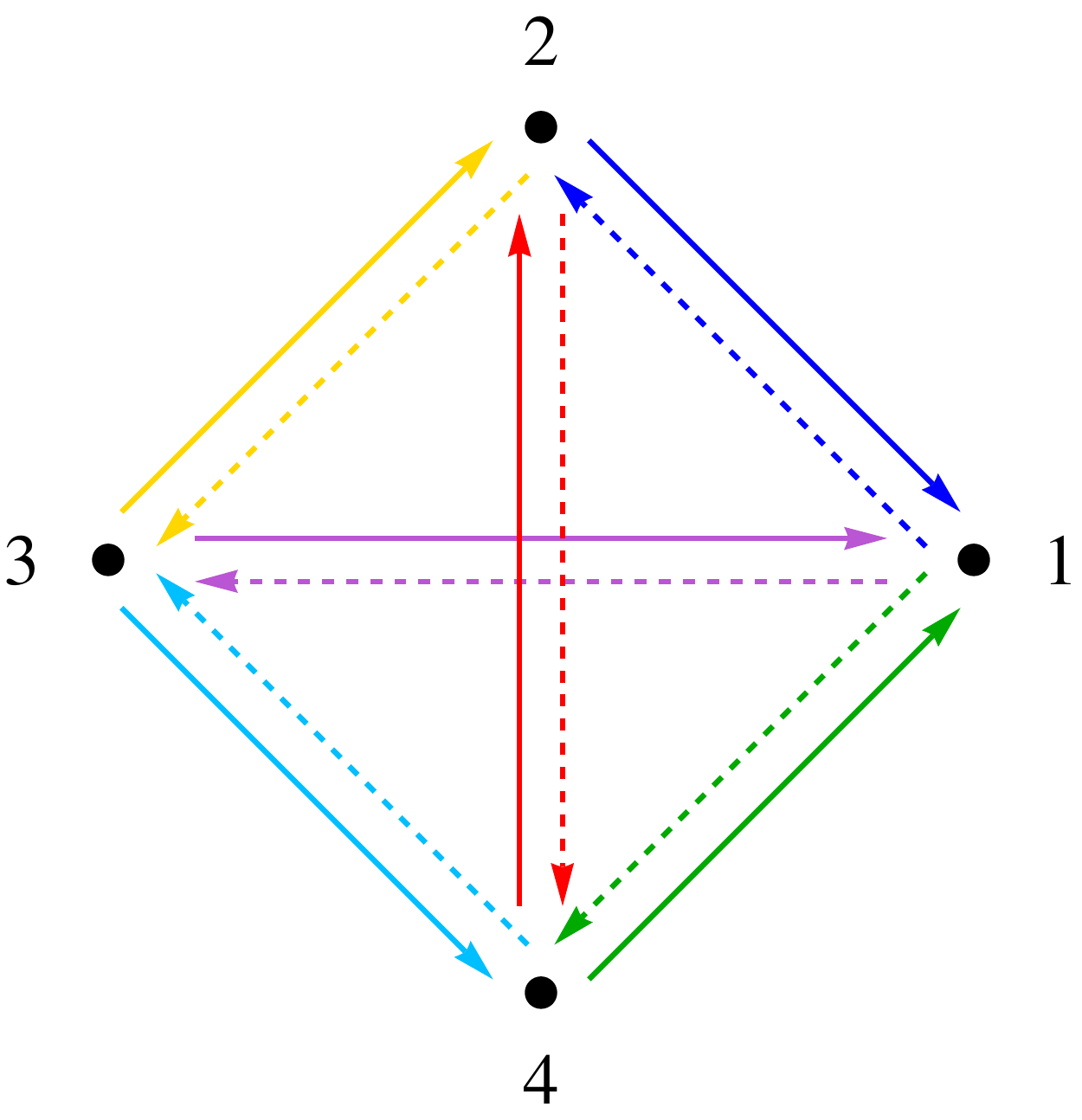}
		\vspace{8pt}
		\caption{on the $v$-plane}
		\label{fig:IRMN4k1_v_plane}
	\end{subfigure}
	\hspace{0em}
	\begin{subfigure}[b]{.3\textwidth}		
		\includegraphics[width=\textwidth]{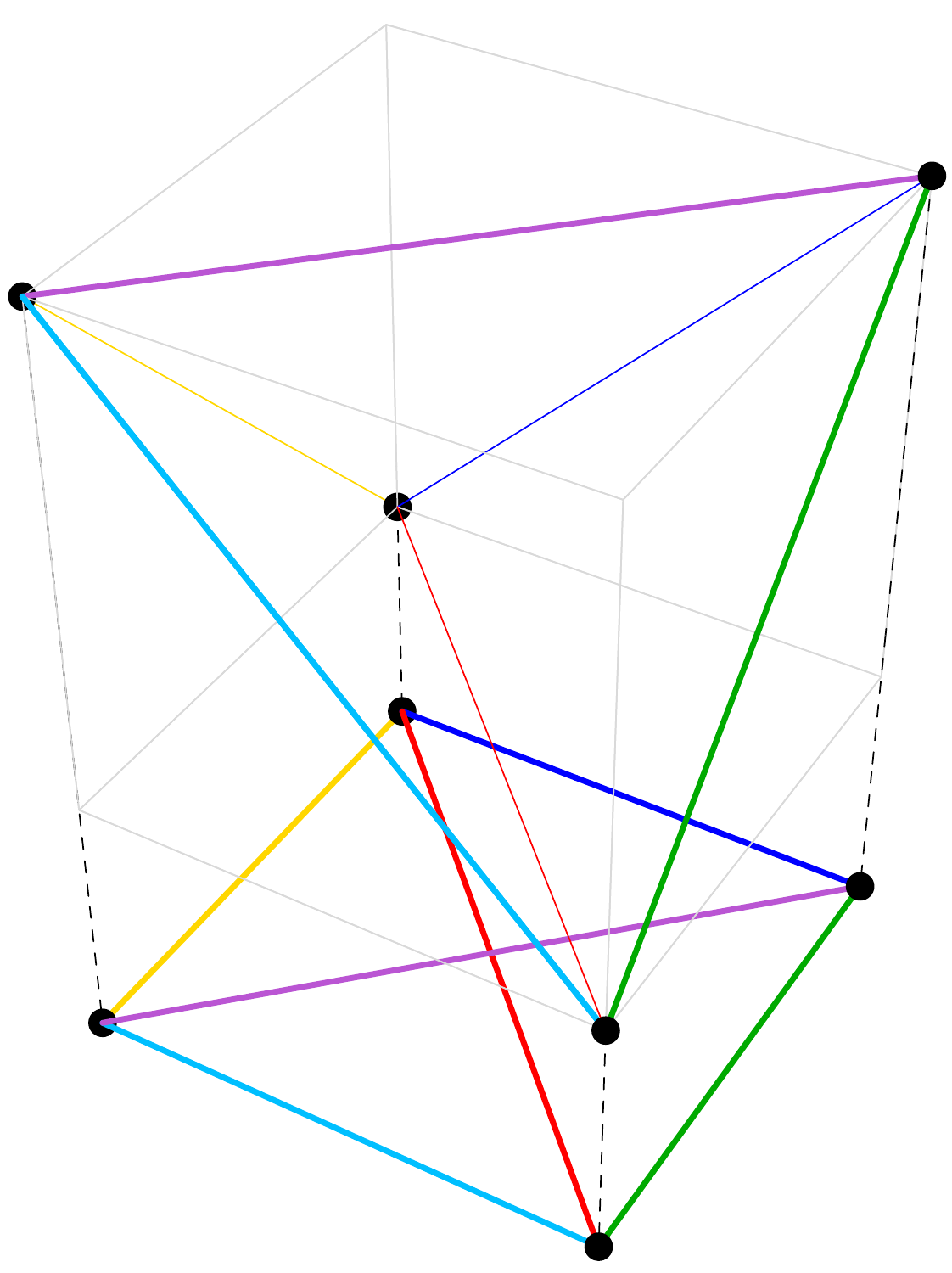}
		\caption{weight space of $A_3$}
		\label{fig:IRMN4k1_weight_space}		
	\end{subfigure}
	\caption{$M'(N=4,k=1,\mu_4)$}
	\label{fig:IRMN4k1}
\end{figure}

Now we consider another example, $N=4$. Figure \ref{fig:IRMN4k1_SWall} shows the spectral network of $M'(N=4,k=1,\mu_4)$. There is a branch point of ramification index 4 from which fifteen $\CS$-walls are coming out, and among them there are 5 pairs of  coincident $\CS$-walls.
From this spectral network we obtain the 2d BPS spectrum of twelve BPS states that can be arranged as shown in Figure \ref{fig:IRMN4k1_v_plane}. This can be understood as the projection of the fundamental representation of $A_3$ and the roots connecting the weights of the representation onto the $W$-plane, which is shown in Figure \ref{fig:IRMN4k1_weight_space}.

\subsection{$M'(N=4,k=2,\mu_4)$}

\begin{figure}[ht]
	\centering
	\begin{subfigure}{.45\textwidth}	
		\includegraphics[width=\textwidth]{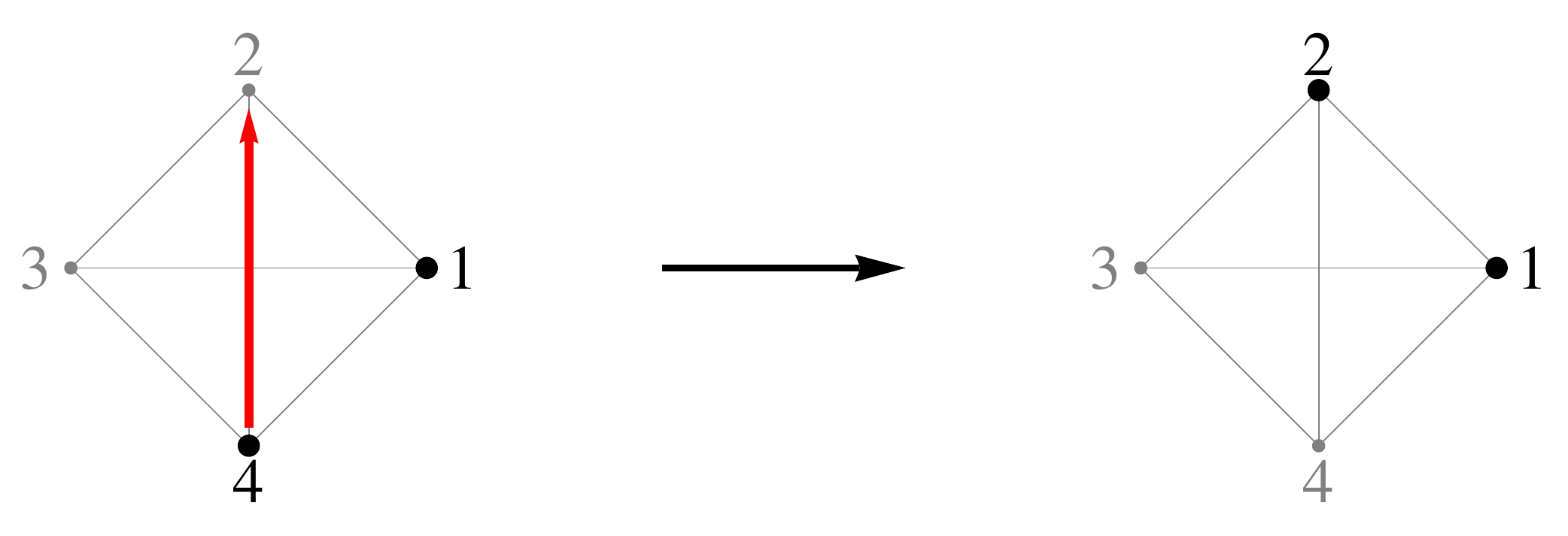}
		\caption{$[14] \to [12]$ soliton for $k=2$}
		\label{fig:IRMN4k2_soliton_diagram_01}
	\end{subfigure}
	\hspace{10pt}
	\begin{subfigure}{.45\textwidth}	
		\includegraphics[width=\textwidth]{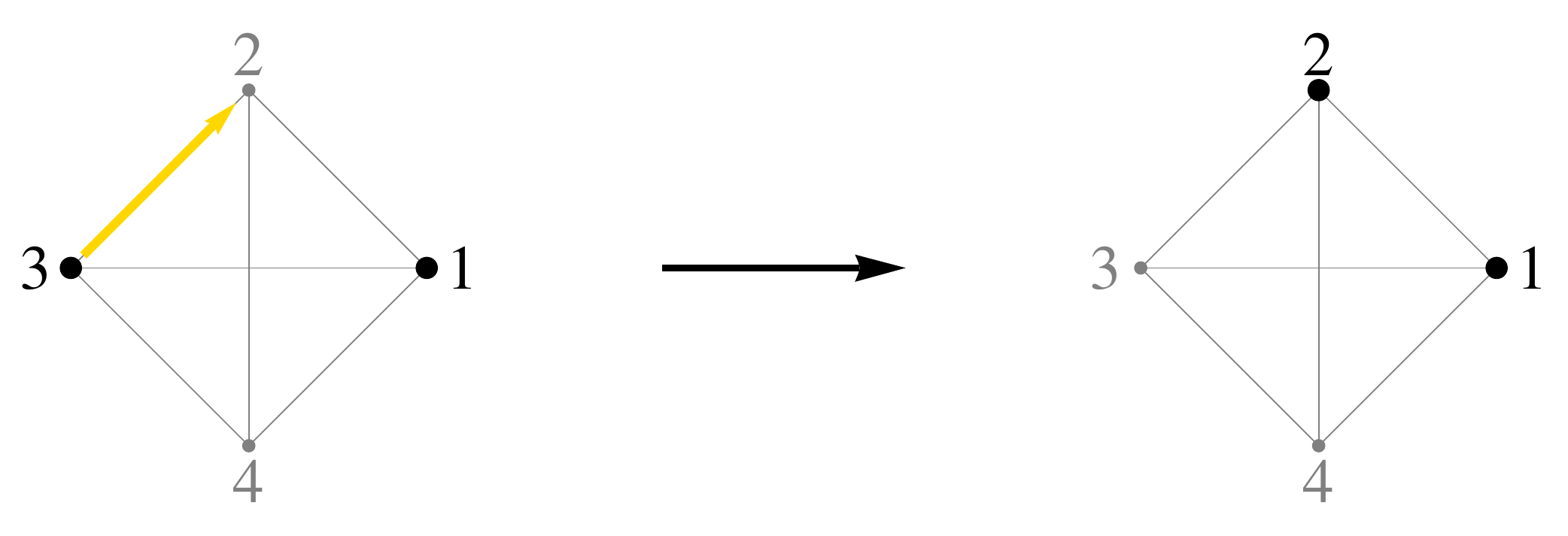}
		\caption{$[13] \to [12]$ soliton for $k=2$}
		\label{fig:IRMN4k2_soliton_diagram_02}
	\end{subfigure}
	\caption{Solitons of $M'(N=4,k=2,\mu_4)$}
	\label{fig:IRMN4k2_soliton}
\end{figure}

Figure \ref{fig:IRMN4k2_soliton} shows that the solitons of $M'(N=4,k=2,\mu_4)$ are again from those of $k=1$.

\begin{figure}[ht]
	\centering
	\begin{subfigure}{.4\textwidth}	
	\centering
		\vspace{20pt}
		\includegraphics[width=\textwidth]{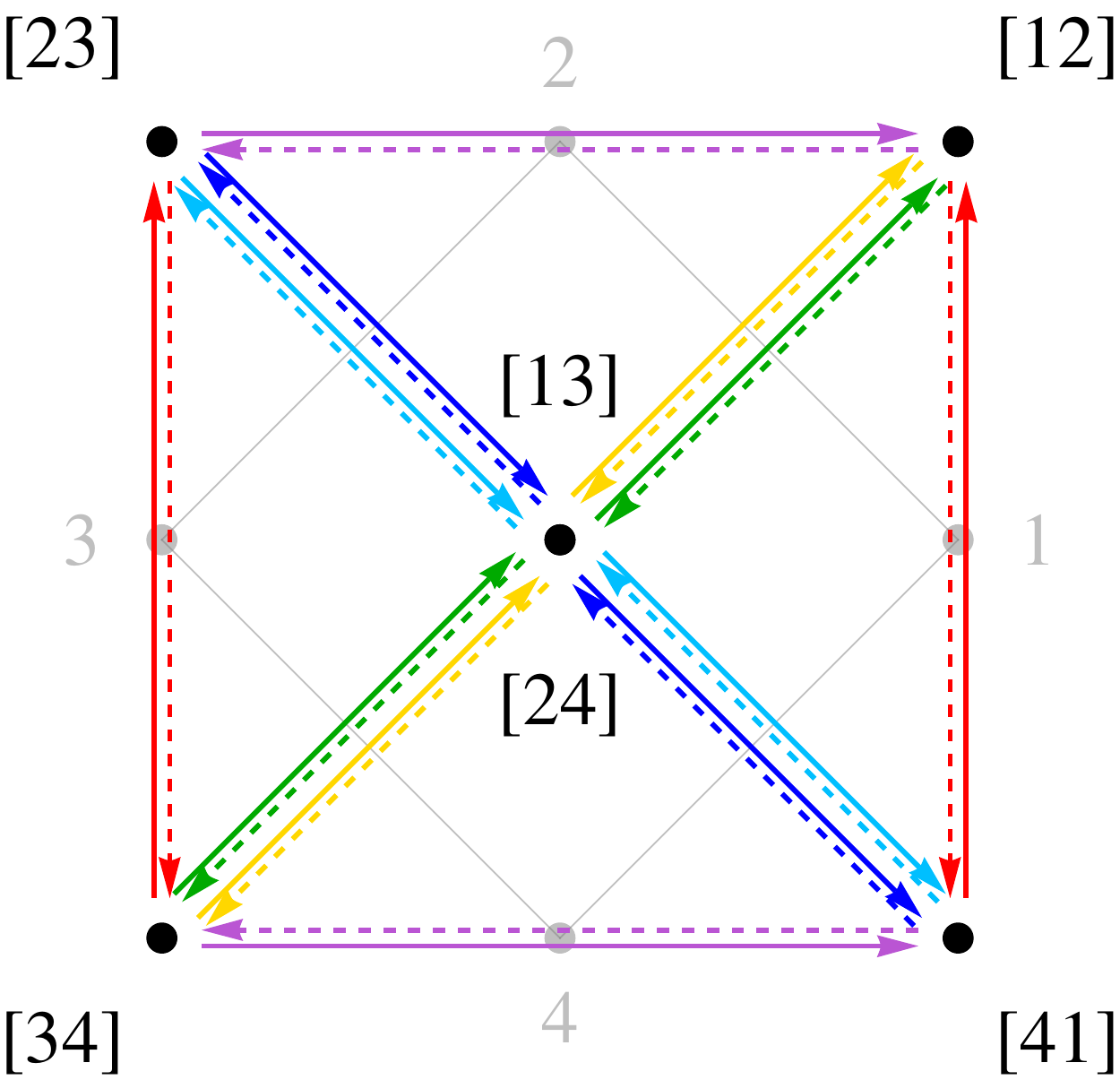}
		\vspace{4pt}
		\caption{on the $v$-plane}
		\label{fig:IRMN4k2_diagram_v_plane}
	\end{subfigure}
	\hspace{.1\textwidth}
	\begin{subfigure}{.4\textwidth}	
		\centering
		\includegraphics[width=\textwidth]{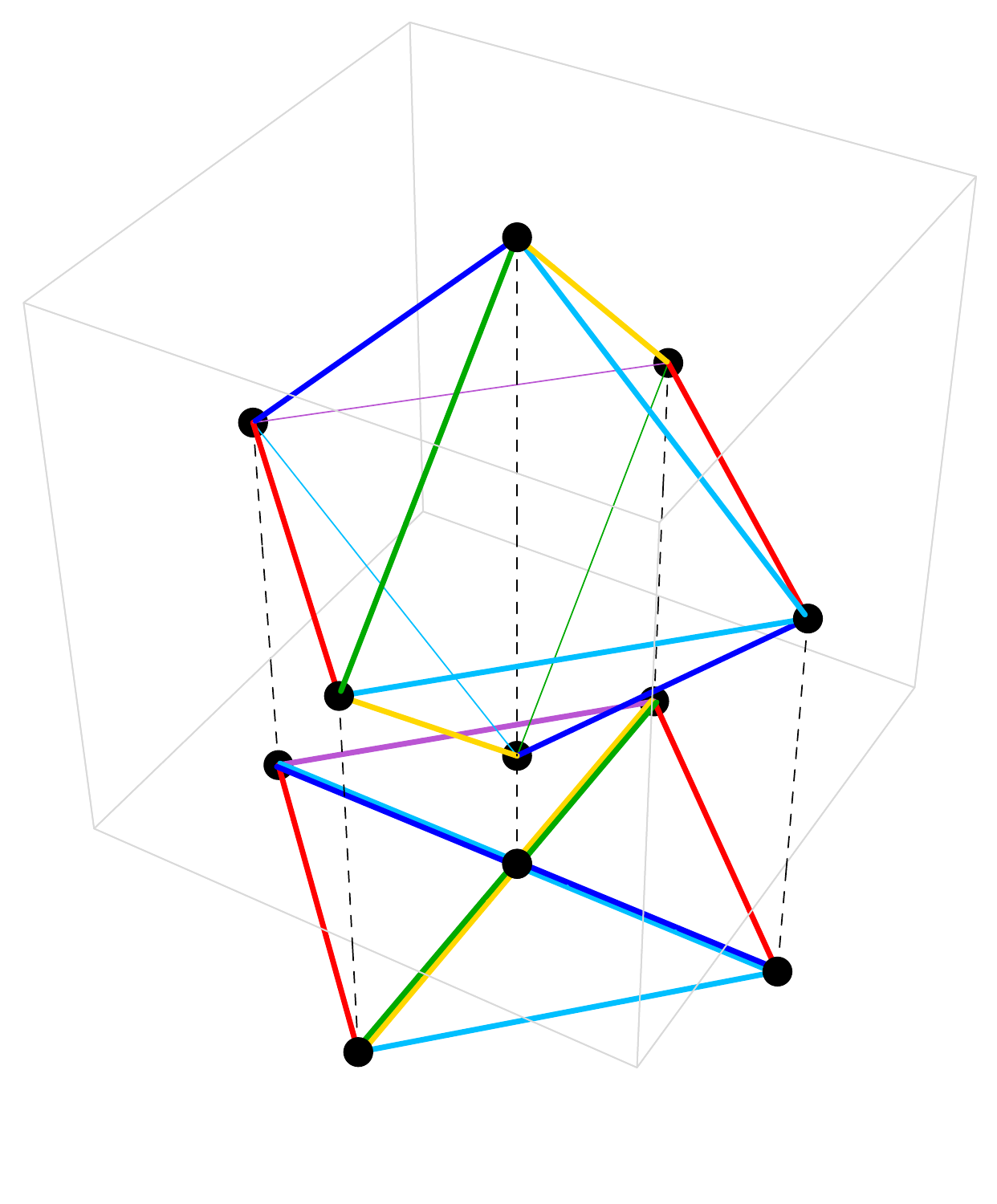}
		\caption{in the weight lattice}
		\label{fig:IRMN4k2_diagram_weight_lattice}
	\end{subfigure}
	\caption{Ground states and solitons of $M'(N=4,k=2,\mu_4)$}
	\label{fig:IRMN4k2_diagram}
\end{figure}

We collect solitons of $M'(N=4,k=2,\mu_4)$ thus obtained in Figure \ref{fig:IRMN4k2_diagram_v_plane}, which is drawn over the $k=1$ BPS spectrum for comparison. 
Again, the result can be understood as projecting the weight space diagram onto the $W$-plane as shown in Figure \ref{fig:IRMN4k2_diagram_weight_lattice}. Note that not every two lines connecting the ground states result in a soliton, but a soliton exists only when the two vacua can be connected by a root of $A_3$. The weight space diagram is symmetric under the reflection with respect to the origin, illustrating $k \leftrightarrow (N-k)$ duality.

\subsection{$M'(N=4,k=3,\mu_4)$}
\begin{figure}[ht]
	\centering
	\begin{subfigure}[b]{.6\textwidth}	
		\includegraphics[width=\textwidth]{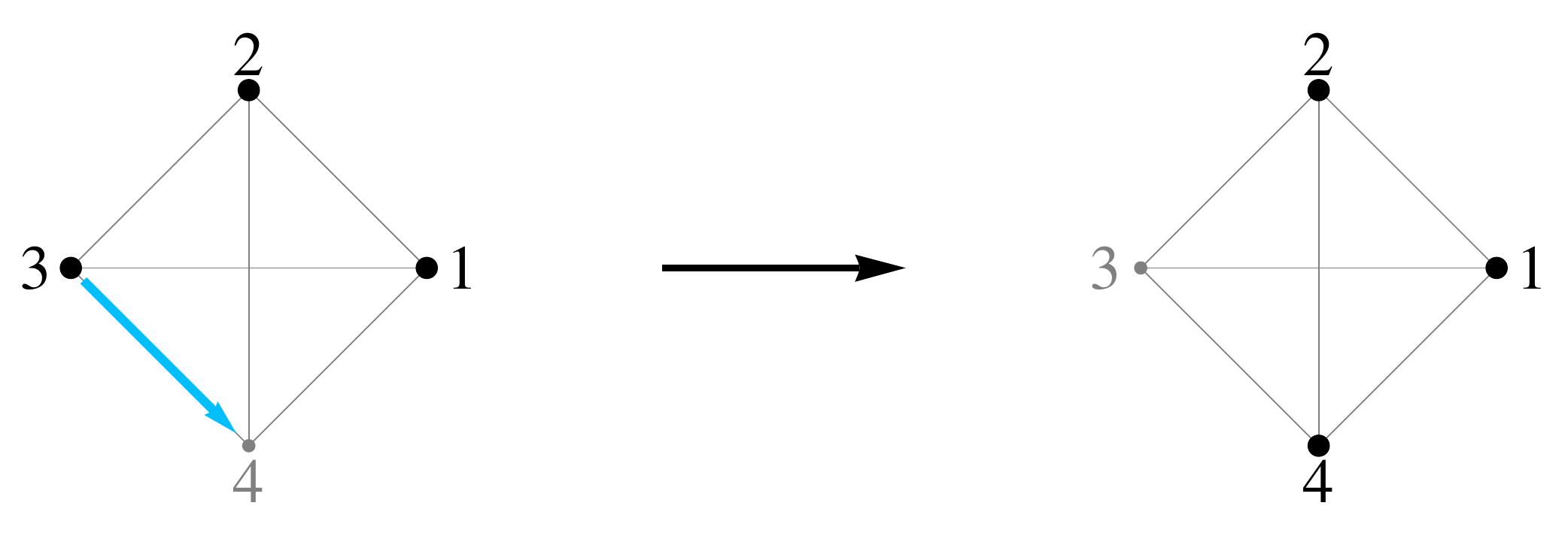}
		\vspace{0pt}
		\caption{$[123] \to [412]$ soliton for $k=3$}
		\label{fig:IRMN4k3_soliton_diagram}
	\end{subfigure}
	\hspace{1em}
	\begin{subfigure}[b]{.3\textwidth}	
	\centering
		\includegraphics[width=\textwidth]{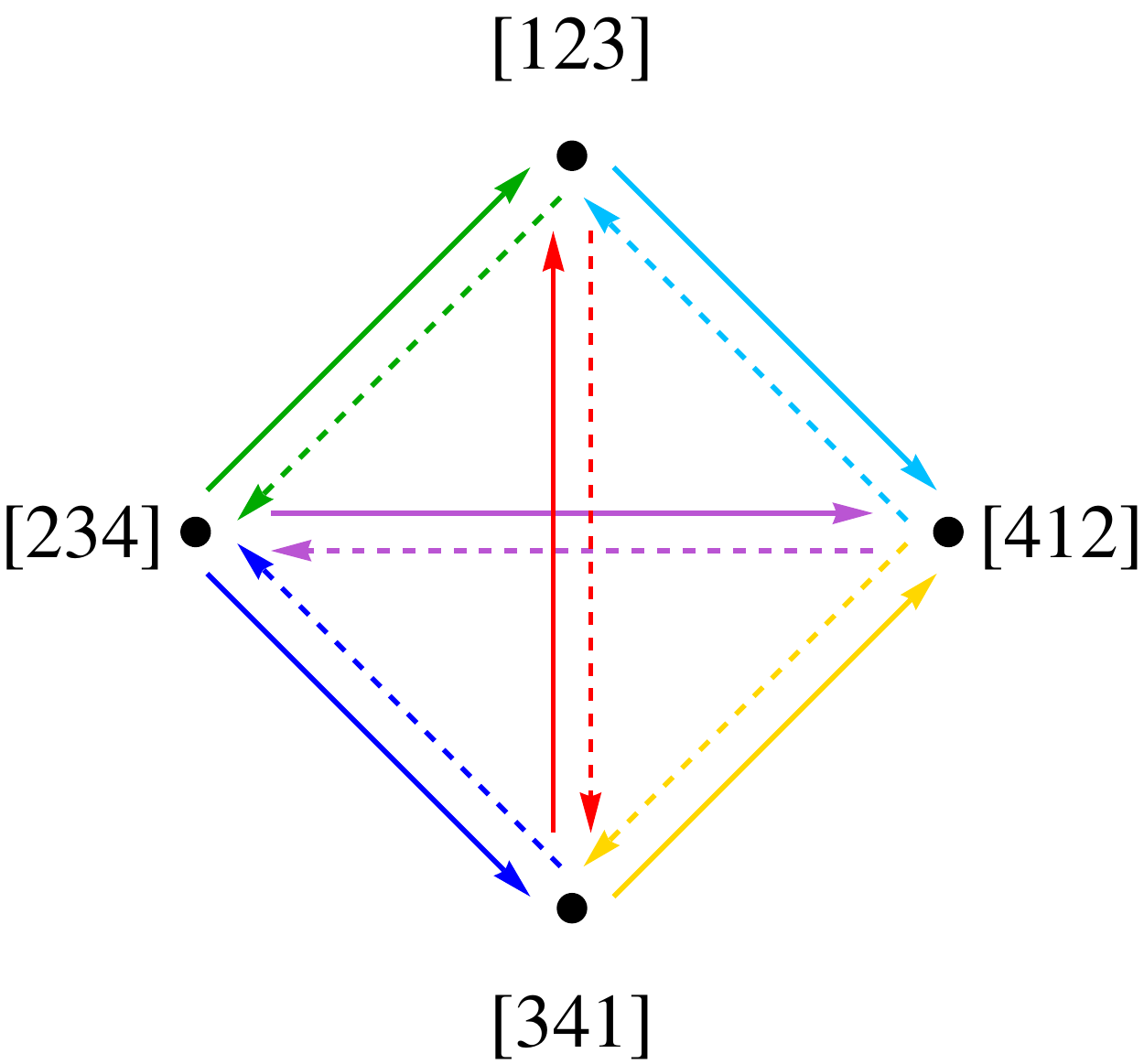}
		\caption{on the $v$-plane}
		\label{fig:IRMN4k3_diagram_v_plane}
	\end{subfigure}
	\caption{Ground states and solitons of $M'(N=4,k=3,\mu_4)$}
	\label{fig:IRMN4k3}
\end{figure}

Figure \ref{fig:IRMN4k3_soliton_diagram} shows how to get a soliton of $M'(N=4,k=3,\mu_4)$ from a $k=1$ soliton. Again, all the solitons of $M'(N=4,k=3,\mu_4)$ are obtained from those of $k=1$, and the BPS spectrum can be represented as Figure \ref{fig:IRMN4k3_diagram_v_plane}. Note that this is isomorphic to the BPS spectrum of $k=1$, thereby showing the $k \leftrightarrow (N-k)$ duality.

\section{Discussion and outlook}
In \cite{Hori:2013ewa} it is proposed that a 2d $\UU(k)$ gauge theory without matter field and with the tree level twisted superpotential
\begin{align*}
	\CW=\tr \Sigma^{N+1}+\pi i(k+1)\tr\Sigma,
\end{align*}
where $\Sigma$ is the field strength for the $\UU(k)$ vector multiplet, flows to the same IR fixed point as the other 2d theories studied here. Evidence is provided by studying its ground states and showing that its chiral ring and $S^2$ partition function agree with those of the Landau-Ginzburg model $LG(N,k)$.

Here we investigated coset models $G/H$ at level $1$ with $G = \SU(N)$, however there are coset models at level greater than 1 and/or involving other classical groups \cite{Kazama:1988qp,Kazama:1988uz}. It would be interesting to find the corresponding brane configuration, which will be helpful in understanding spectral networks in various representations of general Lie group.

The same coset model appears as a dual CFT of the supersymmetric higher-spin theories in AdS$_3$ \cite{Creutzig:2011fe,Gaberdiel:2010pz}. It will be interesting to find applications of the result of \cite{Hori:2013ewa} in this context.

\bibliographystyle{amsplain}
\bibliography{ref}

\end{document}